\documentclass{article}
\usepackage[utf8]{inputenc}\DeclareUnicodeCharacter{2212}{-}
\usepackage{pgf}
\usepackage{import}
\usepackage[normalem]{ulem}

\usepackage[left=2.5cm,right=2.5cm]{geometry}

\newcommand{\be}{\begin{equation}}
\newcommand{\ee}{\end{equation}}
\newcommand{\bea}{\begin{eqnarray}}
\newcommand{\eea}{\end{eqnarray}}

\newcommand{\Tr}{{\rm Tr}}

\renewcommand{\Re}{\mathrm{Re} \,}

\def\cyi{a}
\def\upi{b}
\def\Oxf{e}
\def\Cut{d}
\def\wup{c}
\usepackage{array,multirow,hhline}
\usepackage{amsmath}
\usepackage{amssymb}
\usepackage{siunitx}
\usepackage{dsfont}
\usepackage{slashed}
\usepackage{mathtools}
\usepackage{braket}
\usepackage{float}

\usepackage{caption}
\usepackage[sorting=none,backend=bibtex,style=numeric-comp,natbib=true]{biblatex} 
\begin{document}

\begin{titlepage}
  \begin{center}
    \begin{LARGE}
      \textbf{Glueball Spectrum with four light dynamical fermions} \\
    \end{LARGE}
  \end{center}

\vspace{.5cm}

\vspace{-0.8cm}
  \baselineskip 20pt plus 2pt minus 2pt
  \begin{center}
    \textbf{A.~Athenodorou\(^{(\cyi, \upi)}\), J.~Finkenrath\(^{(\wup, \cyi)}\),
    A.~Lantos\(^{(\Cut)}\),
    M.~Teper\(^{(\Oxf)}\)
    }
  \end{center}
  
  \begin{center}
    \begin{footnotesize}
        \noindent
        \(^{(\cyi)}\) Computation-based Science and Technology Research Center, The Cyprus Institute, 20 Kavafi Str., Nicosia 2121, Cyprus \\
        \(^{(\upi)}\) Dipartimento di Fisica, Università di Pisa and INFN, Sezione di Pisa, Largo Pontecorvo 3, 56127 Pisa, Italy \\
         \(^{(\wup)}\) Bergische Universit\"at Wuppertal, Gau{\ss}stra{\ss}e 20, Wuppertal, Germany. \\
        \(^{(\Cut)}\) Cyprus University of Technology, Archiepiskopou Kyprianou 30, Limassol 3036, Cyprus\\
       \(^{(\Oxf)}\) All Souls College and Department of Physics, University of Oxford, Oxford OX1 2JD, United Kingdom\\
    \end{footnotesize}
  \end{center}

  \begin{abstract} 
      {\color{black} We perform a calculation of the glueball spectrum for $N_f=4$ degenerate dynamical fermions with masses corresponding to light pions. We do so by making use of ensembles produced within the framework of maximally twisted fermions by the Extended Twisted Mass Collaboration (ETMC). We obtain masses of states that fall into the irreducible representations of the octahedral group of rotations in combination with the quantum numbers of charge conjugation $C$ and parity $P$; the above quantum numbers result in 20 distinct irreducible representations. We implement the Generalized Eigenvalue Problem (GEVP) using a basis that consists only of gluonic operators. The purpose of this work is to investigate the effect of light dynamical quarks on the glueball spectrum and how this compares to the statistically more accurate spectrum of $SU(3)$ pure gauge theory. Given that glueball states may have broad widths and thus need to be disentangled from all the relevant mixings, we use large ensembles of the order of ${\sim {~\cal O}}(20 {\rm K})$ configurations. Despite the large ensembles, the statistical uncertainties allow us to extract the masses for only a few irreducible representations; namely $A_1^{++}$, $A_1^{-+}$, $E^{++}$ as well as $T_2^{++}$. The results for the scalar $A_1^{++}$ representation show that an additional state appears as the lightest state in the scalar $A_1^{++}$  channel of the glueball spectrum, while the next two excited states are consistent with the lightest two states of the pure gauge theory. To further elucidate the nature of this additional state we perform a calculation using $N_f=2+1+1$ configurations and this demonstrates that it possesses a large quark content. Finally, the ground states of the $E^{++}$ and $T_2^{++}$ tensor channels and of the $A_1^{-+}$ pseudoscalar channel show, at most, minor effects due to the inclusion of dynamical quarks.} 

  \begin{center}                                                                             \today
  \end{center}
  \end{abstract}
\end{titlepage}                                                                                                                     
{\color{black}
\section{Introduction}
\label{sec:introduction}
The extraction of the spectrum of glueballs in full QCD at physical quark masses and in the continuum limit is an open question which requires a coordinated effort from the Lattice community. A number of experiments such as PANDA~\cite{Parganlija:2013xsa} and BESIII~\cite{Asner:2008nq} are currently looking to identify such states, and so the extraction of the associated spectrum from first principles as well as the understanding of how light dynamical quarks affect the spectrum of glueballs is a timely question that merits careful investigation. Recent reviews on searches for glueballs can be found in the Lattice 2022 plenary presentation by Davide Vadacchino~\cite{vadacchino_davide_2022_7338133,Vadacchino:2023vnc} as well as in the review in Ref.~\cite{Klempt:2022ipu} by Eberhard Klempt. Glueball states have large fluctuations and possibly broad resonance widths and so the precise determination of their spectrum requires high statistics, at least of the order of ${{\cal O}} (10-100{\rm K})$ configurations~\cite{Gregory:2012hu}. Hadron physics performed using current Lattice methodology and machinery, on the other hand, makes use of sets of typically a few hundred configurations. Nevertheless, for a number of measurements on the lattice such as the extraction of renormalisation factors, large sets of configurations are produced and can thus be used for exploratory investigations of the glueball spectrum. This work results from the availability of accessing such large sets of configurations.  

In this work we invested effort in the investigation of the effect of light dynamical quarks on the spectrum of glueballs. The spectrum of QCD in the presence of light quarks is of interest for both phenomenological and theoretical reasons. On the phenomenological side it would provide important information regarding the mixing of ${q {\bar q}}$ states with glueball states. This is expected to be modest by the OZI rule~\cite{OKUBO1963165}, which could mean that, in practice, states with quark content could actually be undetectable in the glueball spectrum, as calculated using purely gluonic correlators, or they might appear with extremely small overlaps. For this purpose we use configurations produced with $N_f = 4$ light twisted mass quarks, since we expect that such a set up will enhance the impact of light dynamical quarks on the spectrum of glueballs.

We extract the glueball spectrum and then compare it with the one calculated using pure gauge $SU(3)$ configurations~\cite{Athenodorou:2020ani}. Using decoupling arguments, we expect that for massive dynamical quarks the glueball spectrum becomes similar to that of the pure gauge theory. The important question which arises here is what happens if one includes dynamical fermions with low masses close to physical values.

Our investigation within $N_f=4$ QCD with four light degenerate fermions reveals the presence in the scalar channel ($R^{PC} = A_1^{++}$) of an extra state which is also the lightest state, in addition to excited states that match the lightest and first excited states in the pure gauge theory. It is possible that this additional state involves some mixing with $q {\bar q}$ states. To understand what is occurring in the low-lying spectrum of the scalar channel and whether this additional state does indeed include a large $q {\bar q}$ component we turn to some available $N_f=2+1+1$ ensembles. Within this setup we extract the ground state energy for the scalar channel and find that, in contrast to the first excited state, this indeed exhibits a strong dependence on the pion mass.

Overall, our main findings can be summarized in the following points.
\begin{enumerate}
    \item In the scalar channel we obtain an additional state when we introduce light dynamical quarks; this is the lightest state with the next two states having masses consistent with the lightest two glueballs in the pure gauge theory.
    \item In the additional calculations that we perform using the $N_f=2+1+1$ gauge ensembles, we observe that the additional ground state in the scalar channel depends strongly on the pion mass, unlike the first excited state. 
    \item  The tensor glueball mass appears to be insensitive to the presence of light dynamical quarks.
    \item The pseudoscalar glueball mass is affected only slightly by the inclusion of light dynamical quarks and is close to the mass of the tensor glueball. 
    \item The string tension is suppressed by the inclusion of four light dynamical quarks by 50-70\%.
\end{enumerate}

This article is organized as follows: In Section~\ref{sec:lattice_set_up} we provide a detailed presentation of the lattice set up which has been used for the production of the configurations with $N_f=4$ and $N_f=2+1+1$ twisted mass fermions as well as those with the $SU(3)$ pure gauge action. Then, in Section~\ref{sec:calculation_of_glueball_masses} we explain how one can extract the spectrum of glueballs in Lattice QCD by making use of the Generalized Eigenvalue Problem. In the same manner, in Section~\ref{sec:torelon_masses} we describe how to evaluate the masses of torelons. Following that, in Section~\ref{sec:topological_charge_and_scale_setting}, we describe the calculation of the topological charge which is used as a measure of the ergodicity of the system. Furthermore we explain how we evaluate the energy scale $t_0$ using the gradient flow. Subsequently in Section~\ref{sec:results}, we move to the presentation of the results by focussing on the scalar glueballs in the $R^{PC}=A_1^{++}$ channel, the lightest tensor glueball whose components are split between the $R^{PC}=E^{++}$ and the $T_2^{++}$ channels, as well as the pseudoscalar glueball obtained in the $R^{PC}=A_1^{-+}$ channel. Then, in Section~\ref{sec:Nf211} we focus our discussion on the ground state obtained in the scalar channel for $N_f=2+1+1$, where it appears to display a dominant $q {\bar q}$ component. Subsequently, in Section~\ref{sec:torelon}, we present our results for the string tension obtained from the energies of the torelons. We then compare our glueball results with other older as well as recent results from the literature, before presenting our conclusions in Section~\ref{sec:conclusions}. }

{\color{black}
\section{Lattice Set-up}
\label{sec:lattice_set_up}

For the evaluation of the glueball spectrum we use gauge ensembles of clover improved twisted mass fermions produced  with \(4\) degenerate light flavours ($N_f=4$) at two different lattice spacings as well as two ensembles with \(2\) degenerate light flavours and a strange and a charm quark ($N_f=2+1+1$). All the configurations with dynamical quarks have been generated within the context of the Extended Twisted Mass collaboration. They were also used in a parallel work for extracting renormalisation constants~\cite{Alexandrou:2020sml}.

For the gluonic part of the action we use the Iwasaki improved gauge action~\cite{Iwasaki:1984cj,Iwasaki:1983gx,Weisz:1982zw}, which is given by the expression
\be
S_G = \frac{\beta}{3} \sum_x \left(c_0\sum_{ \substack{\mu,\nu=1 \\ \mu<\nu} }^4\left[ 1- \Re\Tr\left( U_{x,\mu\nu}^{1\times1}\right)\right] + c_1 \sum_{ \substack{\mu,\nu=1 \\ \mu\neq\nu} }^4 \left[ 1- \Re\Tr\left(U_{x,\mu\nu}^{1\times2}\right)\right]\right)\,,
\label{eq:g_action}
\ee
where \(\beta=6/g^2\), \(U^{1\times1}\) is a plaquette and \(U^{1\times2}\) is a rectangular Wilson loop. The Symanzik coefficients are \(c_0=3.648\) and \(c_1= (1-c_0)/8\). The fermionic action is given
 \cite{Frezzotti:2000nk,Frezzotti:2003ni} by:
\be
S_F^{l} = a^4\sum_x \bar{\chi}^{(l)}(x)\left( D_W[U] + \frac{i}{4} c_{SW}\sigma^{\mu\nu}\mathcal{F}^{\mu\nu}[U] + m_{0,l} + i \mu_l\gamma_5\tau^3 \right)\chi^{(l)}(x)\,.
\label{eq:fl_action}
\ee
In the equation above, \(\chi^{(l)}\) is the field representing the quark doublets, expressed in the twisted basis, \(m_{0,l}\) and \(\mu_l\) are respectively the untwisted and twisted  mass parameters, \(\tau^3\) is the third Pauli matrix acting in flavor space and \(D_W\) is the massless Wilson-Dirac operator. The clover term \(\propto \sigma^{\mu\nu}\mathcal{F}^{\mu\nu}\) is included in the action to suppress cut-off effects, reducing the difference between the mass of the charged and neutral pions~\cite{Alexandrou:2018egz}.

Moving on to the $N_f=2+1+1$ case, the additional strange and charm quarks are included as a non-degenerate twisted doublet \(\chi^{(h)}=(s,c)^t\), with the  action~\cite{Frezzotti:2003xj}
\be
S_F^{h} = a^4\sum_x \bar{\chi}^{(h)}(x)\left( D_W[U] + \frac{i}{4} c_{SW}\sigma^{\mu\nu}\mathcal{F}^{\mu\nu}[U] + m_{0,h} - \mu_{\delta}\tau^1 + i \mu_{\sigma}\gamma_5\tau^3 \right)\chi^{(h)}(x)\,,
\label{eq:fh_action}
\ee
 where \(m_{0,h}\) is the bare untwisted quark mass for the heavy doublet, \(\mu_{\delta}\) the bare twisted mass along the \(\tau^1\) direction and \(\mu_{\sigma}\) the mass splitting in the \(\tau^3\) direction.

The partial conserved axial current (PCAC) mass is tuned to zero in order to achieve maximal twist. This ensures automatic \(\mathcal{O}(a)\) improvement for the expectation values of the observables of interest~\cite{Frezzotti:2005gi}. The simulation parameters of the gauge ensembles with $N_f=4$ as well as $N_f=2+1+1$ quarks are given in Tables~\ref{tab:params_sim_Nf4} and \ref{tab:params_sim_Nf2+1+1} respectively.



\begin{table}[H]
    \centering
    \begin{tabular}{l|c|l|c|c|c|c}
        \hline \hline
  & $\beta$ & $\quad c_{SW}$ & $\mu_l$ & $L$ & $a m_{PS}$ & $t_0/a^2$ \\
        \hline \hline
\texttt{cB4.06.16} & $1.778$ & $1.69$ & $0.006$ & 16 & $0.2652(53)$ & $4.947(62)$  \\
\texttt{cB4.06.24} & $1.778$ & $1.69$ & $0.006$ & 24 & $0.1580(\phantom{0}8)$ & $4.667(17)$  \\
\texttt{cC4.05.24} & $1.836$ & $1.6452$ & $0.005$ & 24 & $0.1546(20)$ & $6.422(48)$  \\
\hline
\end{tabular}
\captionof{table}{ Simulation parameters of the $N_f=4$ gauge ensembles~\cite{Alexandrou:2020sml} used in this work.}
    \label{tab:params_sim_Nf4}
\end{table}
\begin{table}[H]
    \vspace{-0.5cm}
    \centering
    \begin{tabular}{l|c|c|c|c|c|c}
        \hline \hline
  & $\beta$ & $\quad c_{SW}$ & $\mu_l$ & $L$ & $a m_{PS}$ & $t_0/a^2$ \\
        \hline \hline
\texttt{cA211.53.24} & $1.726$ & $1.74$ & $0.0053$ & 24 & $0.1661(\phantom{0}4)$ & $2.342(\phantom{0}6)$  \\
\texttt{cA211.25.32} & $1.726$ & $1.74$ & $0.0025$ & 32 & $0.1253(\phantom{0}1)$ & $2.392(\phantom{0}4)$  \\
\hline
\end{tabular}
\captionof{table}{ Simulation parameters of the $N_f=2+1+1$ gauge ensembles ~\cite{Alexandrou:2018egz,Alexandrou:2018sjm} used in this work.}
    \label{tab:params_sim_Nf2+1+1}
    \vspace{-0.25cm}
\end{table}
For the purpose of comparison with the $SU(3)$ pure gauge theory, we also simulate that theory using the standard Wilson action which is given by Eq.~\ref{eq:g_action} with Symanzik coefficients $c_0=1$ and $c_1=0$. The simulation algorithm combines standard heat-bath and over-relaxation steps in the ratio 1:4; these are implemented by updating $SU(2)$ subgroups using the Cabibbo-Marinari algorithm~\cite{Cabibbo:1982zn}. The parameters of the pure gauge runs are shown in Table~\ref{tab:params_sim_pure_gauge}.
\begin{table}[H]
    \centering
    \begin{tabular}{c|c|c|c|c}
        \hline \hline
 $\beta$ & $L$ & $\sqrt{\sigma}$ & $a m_{G}$ & $t_0/a^2$ \\
        \hline \hline
$6.222$ & $30$ & $0.1533(6)$ & $0.499(6)$ & $6.422$ \\    
$6.135$ & $26$ & $0.1750(9)$ & $0.578(4)$  & $4.947$  \\
$6.117$ & $26$ & $0.1781(12)$ & $0.585(8)$ & $4.667$  \\
\hline
\end{tabular}
\captionof{table}{Simulation parameters for the $SU(3)$ pure gauge ensembles used in this work. The last column represents the values of $t_0/a^2$ for $N_f=4$ simulations corresponding to the given values of $\beta$ via a cubic spline interpolation of data from Refs.~\cite{Luscher:2010iy,Francis:2015lha,Luscher:2011kk,Ce:2015qha} and presented in Figure~\ref{fig:t_0_beta_interpolation}.}
    \label{tab:params_sim_pure_gauge}
\end{table}

\begin{figure}[h]
    \centering
    \includegraphics[height=8cm]{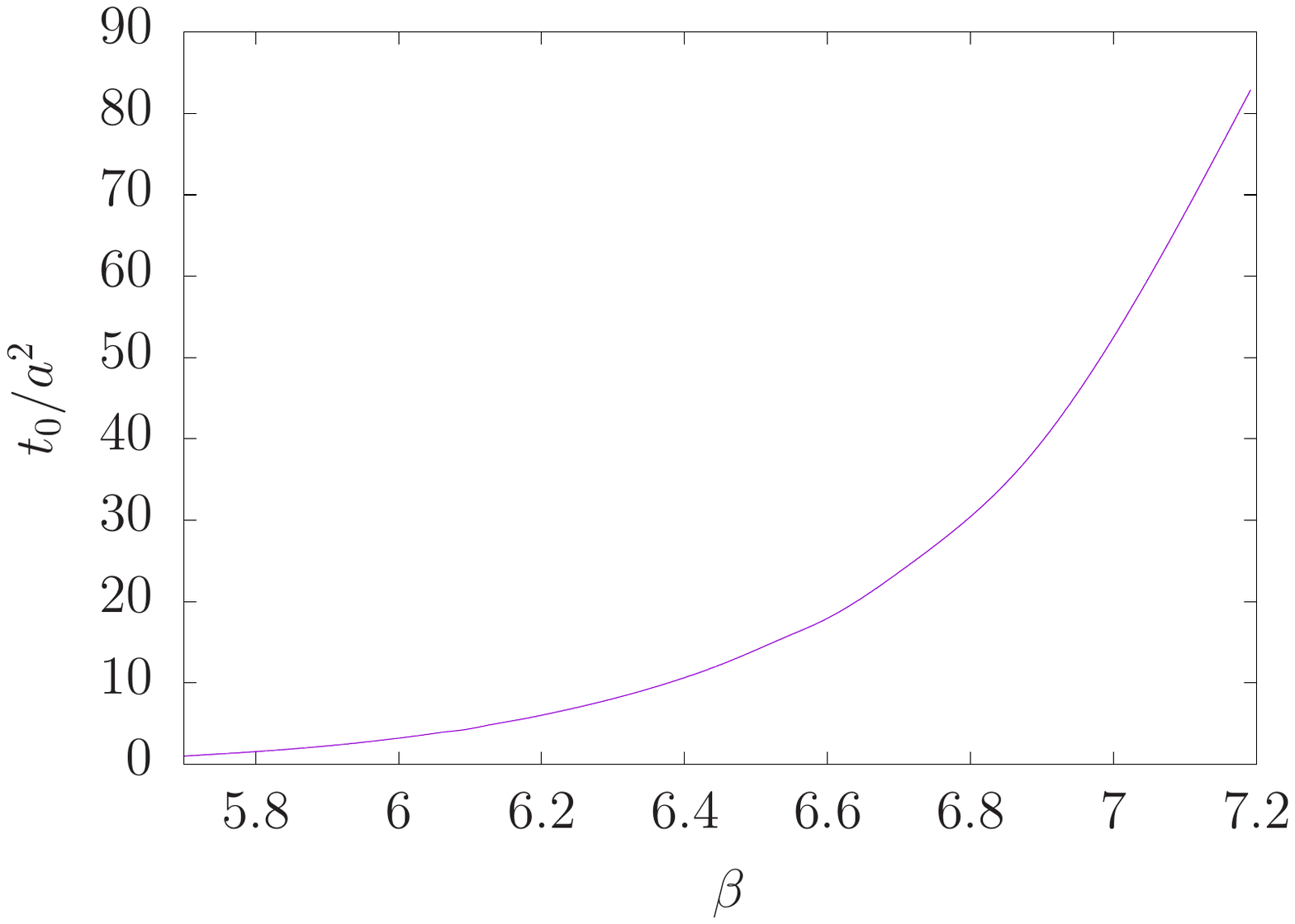}
    \caption{The cubic spline interpolation between $t_0/a^2$ and $\beta$ extracted from results taken from Refs~\cite{Luscher:2010iy,Francis:2015lha,Luscher:2011kk,Ce:2015qha}.}
    \label{fig:t_0_beta_interpolation}
\end{figure}

The idea behind choosing the particular parameters is based on matching the gradient flow times $t_0/a^2$ between the $N_f=4$ QCD and pure gauge runs. Since $N_f=4$ QCD configurations had already been produced, we calculate the corresponding values for $t_0/a^2$ and subsequently match them to the right value of $\beta$ for the pure gauge theory using cubic spline interpolations based on data from Refs~\cite{Luscher:2010iy,Francis:2015lha,Luscher:2011kk,Ce:2015qha}. The curve obtained from the cubic spline interpolation is given in Figure~\ref{fig:t_0_beta_interpolation}. The motivation for this procedure is discussed in more depth in Section~\ref{sec:results}.}

\newpage

\begin{figure}[h]
    \centering
    \includegraphics[height=9cm]{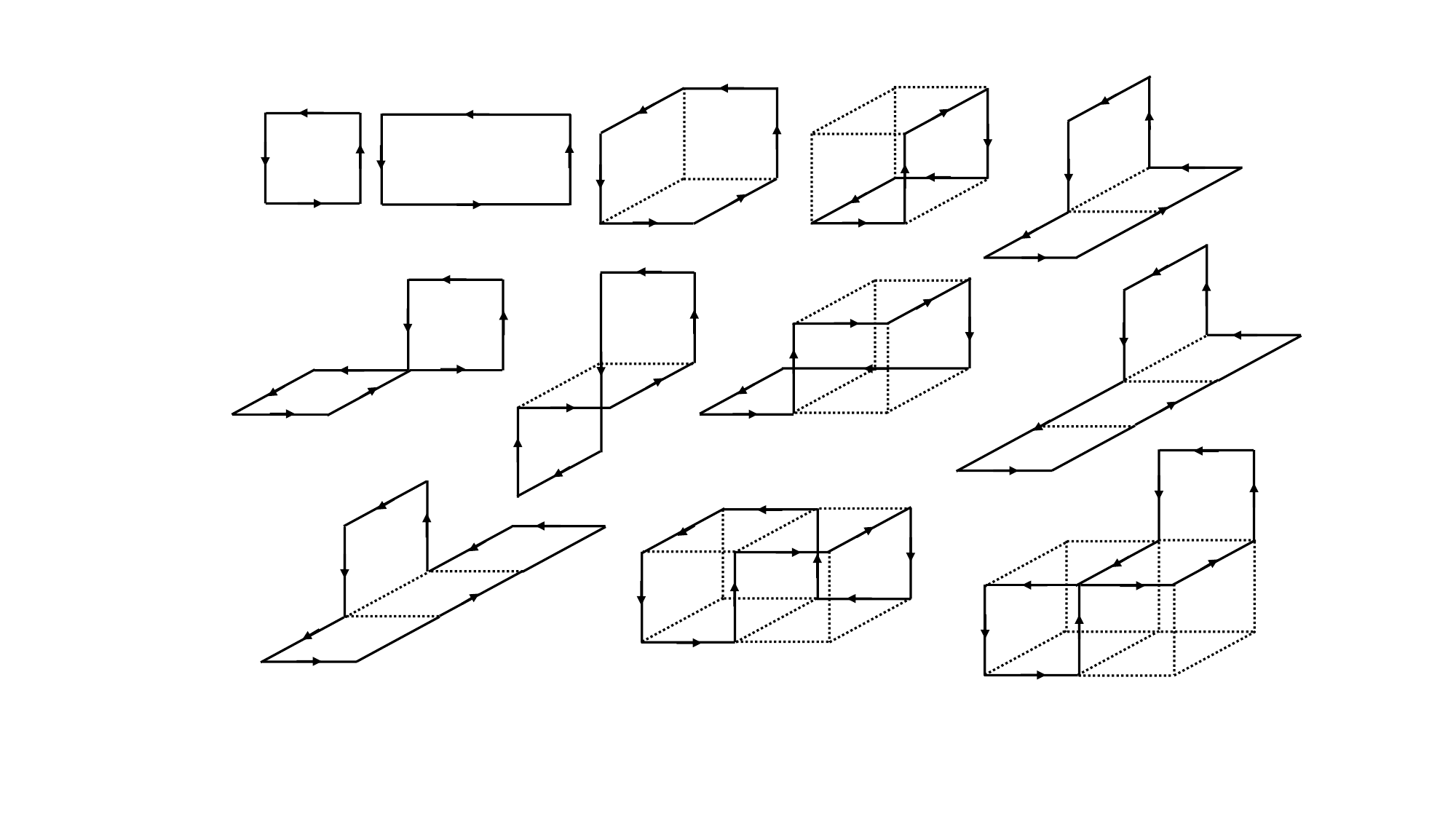}
    \caption{All the different closed loops used for the construction of the glueball operators.}
    \label{fig:glueball_operators}
\end{figure}

{\color{black}
\section{Calculation of Glueball Masses}
\label{sec:calculation_of_glueball_masses}
Glueballs are colour singlet states and their masses can be determined by employing the standard decomposition of a Euclidean correlator associated with an operator denoted as $\phi(t)$. This correlator has an overlap with the physical states expressed in the form of the eigenstates of the system's Hamiltonian, represented as $H$:
\begin{eqnarray}
\langle \phi^\dagger(t=an_t)\phi(0) \rangle
& = &
\langle \phi^\dagger e^{-Han_t} \phi \rangle
=
\sum_i |c_i|^2 e^{-aE_in_t} \nonumber \\
& \stackrel{t\to \infty}{=} & 
|c_0|^2 e^{-aE_0n_t}\,,
\label{extract_mass}
\end{eqnarray}
where the energy levels are arranged in ascending order as $E_{i+1}\geq E_i$, where $E_0$ represents the ground state energy. Only states with non-zero overlaps, denoted as $c_i = \langle {\rm vac} | \phi^\dagger | i \rangle \neq 0$, contribute to the summation above. Thus, it is necessary to ensure that the quantum properties of the operator $\phi$ match those of the states we are interested in. In our study, our primary focus is on glueballs and torelons, so we must embed the appropriate quantum numbers within the operator $\phi$ to project onto these specific states. These quantum numbers include the angular momentum-spin $J$, Parity-$P$, and Charge conjugation-$C$, which define the properties of glueball states.

The determination of the ground state hinges on two key factors: the quality of the overlap with this state and the rapidity of exponential decay with respect to the time variable $t$, as stipulated by Eq.(\ref{extract_mass}). Maximizing the overlap involves constructing operators that effectively encapsulate the essential properties of the target state. Achieving this entails projecting onto the appropriate quantum characteristics and physical length scales associated with the relevant state. To expedite the onset of exponential decay with respect to $t$, it becomes imperative to minimize the influence of higher excited states. To address this, we employ the Generalized Eigenvalue Problem (GEVP)\cite{Luscher:1984is,Luscher:1990ck}, which is applied to a set of operators constructed using various lattice loops at different blocking levels~\cite{Lucini:2004my,Teper:1987wt}. This approach mitigates contamination from excited states, enhancing the operators' alignment with the physical length scales.

The glueballs are color singlets, and thus, an operator designed to represent a glueball state is created by taking the trace of an ordered product of $SU(3)$ link matrices along a contractible loop. To ensure that the correlators maintain exact positivity, we restrict the use of loops to those containing only spatial links. The real part of the trace corresponds to Charge conjugation $C = +$, while the imaginary part corresponds to $C = -$. To generate an operator with zero momentum ($p=0$), we sum over all spatial translations of the loop. By considering all possible rotations of the loop, we construct linear combinations that conform to the irreducible representations $R$ of the rotational symmetry group associated with our cubic spatial lattice. We consistently opt for a cubic spatial lattice volume ($L_x=L_y=L_z$) that respects these symmetries. For each loop, we also create its parity-inverse counterpart, allowing us to form linear combinations and construct operators that encompass both parities, denoted as $P = \pm$. The correlators of these operators will project onto glueballs with $p = 0$ and possess the quantum numbers $R^{P C}$ as defined by the relevant operators. In total, there are 12 paths used in constructing the glueball operators, which are illustrated in Figure~\ref{fig:glueball_operators}.

The irreducible representations denoted as $R$ within our subgroup of the complete rotation group are conventionally labeled as $A_1, A_2, E, T_1,$ and $T_2$. $A_1$ representation is a singlet and exhibits rotational symmetry, thus, containing the $J=0$ state in the continuum limit. Similarly, the $A_2$ representation is also a singlet, while the $E$ representation forms a doublet, and both $T_1$ and $T_2$ are triplets. For instance, when considering the three states that transform under the triplet of $T_2$, they are degenerate on the lattice, so we average their values and treat them as a single state in our calculations of glueball masses. The same approach is applied to the $E$ doublets to maintain consistency in our estimates.

Glueball energy values are extracted through the utilization of correlation matrices denoted as $C_{ij} = \langle \phi_i^{\dagger} (t) \phi_j (0) \rangle$, where $i,j=1...N_{\rm op}$ and $N_{\rm op}$ representing the number of operators. In the scalar channel $A_1^{++}$, there exists a non-zero projection onto the vacuum. In such cases, it can be convenient to employ the vacuum-subtracted operator $\phi_i - \langle \phi_i \rangle$. This subtraction effectively eliminates the vacuum contribution from Equation~\ref{extract_mass}, ensuring that the lightest non-trivial state appearing in the summation becomes the dominant term in the state expansion.

The representations of the rotational symmetry group described above are specific to our cubic lattice formulation. As we approach the continuum limit, these states will converge towards the continuum glueball states, which are categorized into representations of the continuum rotational symmetry group. In simpler terms, they fall into degenerate multiplets of states with $2J + 1$ states each. When determining the continuum limit of the low-lying glueball spectrum, it becomes more meaningful to assign the states to a particular spin value $J$ rather than to the representations of the cubic subgroup, which provide less detailed information, as they map all spins from $J =1,2,3, \dots, \infty$ to just five cubic representations. The distribution of $2J + 1$ states for a given $J$ among the representations of the cubic symmetry subgroup is outlined in Table~\ref{tab:table_J_R} for relevant low values of $J.$ For example, the five states corresponding to a $J = 2$ glueball will be divided between a degenerate doublet $E$ and a degenerate triplet $T_2$, resulting in a total of five states. These $E$ and $T_2$ states will be affected by lattice spacing corrections of the order $O(a^2)$, but as the lattice spacing $a$ becomes sufficiently small, these states will become nearly degenerate. This near-degeneracy can then be utilized to identify the continuum spin of the states.
\begin{table}[ht]
\begin{center}
 \begin{tabular}[h]{|c||ccccc|}
\hline \hline
    $J$ & $A_1$ & $A_2$ & $E$ & $T_1$ & $T_2$ \\
    \hline \hline
    0 & 1 & 0 & 0 & 0 & 0 \\
    1 & 0 & 0 & 0 & 1 & 0 \\
    2 & 0 & 0 & 1 & 0 & 1 \\
    3 & 0 & 1 & 0 & 1 & 1 \\
    4 & 1 & 0 & 1 & 1 & 1 \\
    5 & 0 & 0 & 1 & 2 & 1 \\
    6 & 1 & 1 & 1 & 1 & 2 \\
    7 & 0 & 1 & 1 & 2 & 2 \\
    8 & 1 & 0 & 2 & 2 & 2 \\
\hline
  \end{tabular}
\end{center}
\caption{Subduced representations of continuum spin $J \downarrow R$ of the octahedral group of rotations up to $J=8$, illustrating the spin content of the
   representations $R$ in terms of the continuum $J$.}
\label{tab:table_J_R}
\end{table}
}

{\color{black}
\section{Calculation of torelon masses}
\label{sec:torelon_masses}
In the same manner as in the glueball mass calculation, we can extract the mass of the closed spatial Polyakov loop, also called a torelon, which winds around a spatial compactified direction. The quantum numbers which characterize the torelon states are the angular momentum-spin $J$, Transverse Parity-$P_{\perp}$ as well as the Longitudinal Parity-$P_{||}$. The ground state of the torelon which provides the easiest way to extract the string tension is described by $J^{P_{\perp}, P_{||}}$=$0^{++}$. For the purposes of this work we will focus on the $0^{++}$channel, leaving the extraction of the torelon mass for several configurations of the quantum numbers  $J^{P_{\perp}, P_{||}}$ a subject of future investigations.

We obtain the mass of the torelon by calculating the ground state energy $m_T(L)$ of a flux-tube of length $L$ that closes on itself by winding once around the spatial compactified torus. We use Eq.~(\ref{extract_mass}) where the operator $\phi$ is the product of $SU(3)$ link matrices taken around a non-contractible closed path that winds once around the spatial torus. The simplest such operator is the elementary Polyakov loop:
\begin{equation}
\phi(n_t) = P(n_t) =  \sum_{n_y,n_z} \mathrm{Tr}
\left\{\prod^{L_x}_{n_x=1} U_x(n_x,n_y,n_z,n_t)\right\} \,,
\label{eqn_poly}
\end{equation}
and, thus, has $0^{++}$ quantum numbers.
                     
The above formula denotes the path ordered product of link matrices in the $x$-direction winding once around the $x$-torus. Then we sum over translations along the $x$-torus and a time slice so that we
project onto zero longitudinal as well as transverse momentum respectively i.e. $(p_x,p_y,p_z) = (0,0,0)$. Confining torelons, at least in pure gauge theories, behave, to an adequate extent, as bosonic strings~\cite{Athenodorou:2010cs}. So far no such observation has been reported in full QCD calculations for the torelon. One would expect that the closed flux-tube behaves as a confining bosonic string, the spectrum of which can be approximated to an adequate accuracy by the Nambu-Goto string~\cite{GODDARD1973109}. The ground state energy of the bosonic string is given by the expression
\begin{eqnarray}
    E(l) = \sigma l \sqrt{1 - \frac{2 \pi}{3 \sigma l^2} }\,,
    \label{eq:nambu_goto}
\end{eqnarray}
where $\sigma$ the string tension and $l=a L_x$ the length of the torelon. By imposing non-zero longitudinal momentum, the energy of the torelon is elevated by a momentum factor as well as by a Kaluza-Klein mode due to the difference between the total contribution of the left and right moving phonons, while by imposing transverse momentum the energy level of the torelon is altered according to the associated lattice dispersion behaviour. Hence, for the purposes of this work we focus on zero longitudinal and transverse momenta.                                                                                                                                 

As explained above, the operator of Eq.~\ref{eqn_poly} is invariant under rotations about its main axis and so has angular momentum $J=0$. It is also clearly invariant under both transverse and longitudinal parity transformations. Therefore, this operator is ideal for projecting onto the torelon absolute ground state $0^{++}$. Once more we employ smearing and blocking techniques in order to enhance the projection onto the physical states. In $SU(N)$ pure gauge theories, torelon operators are not invariant under center symmetry transformations and, thus, their vacuum expectation value is zero as long as this is not broken. However, in full Lattice QCD, the action is not invariant under center symmetry transformation and thus the vacuum expectation value should be subtracted from the correlator.

In the presence of a quark-dynamical background, flux-tubes should decay to pions as long as this is energetically favorable. That is to say, if the flux-tube ground energy is equal or larger than twice the mass of the pion, it should decay to two pions leading to a constant ground energy while the linearly rising contribution "jumps up" to the next excitation level. Of course such a phenomenon has never been observed in the closed flux-tube channel and would be extremely interesting to test this scenario.   
}

\newpage
{\color{black}
\section{Topological charge and scale setting}
\label{sec:topological_charge_and_scale_setting}
{\color{black}
In the continuum theory, the topological charge is calculated as the integral across the four-dimensional volume of the topological charge density, specifically:
\be
{\cal Q} = \frac{1}{32\pi^2} \int d^4 x \: \epsilon_{\mu\nu\rho\sigma} \Tr\left[F_{\mu\nu}(x)F_{\rho\sigma}(x)\right] \,.
\label{eq:Q_continuum_def}
\ee
The discrete equivalent of this quantity can be derived by substituting the gluonic field tensor with a lattice operator, ensuring that it converges correctly to the continuum limit. This selection is not exclusive, and operators with reduced discretization effects can be obtained by employing \({\cal O}(a)\)-improved definitions of \(F_{\mu\nu}\). Specifically, the \({\cal Q}\) definition utilized in our study is the symmetric or 'clover' definition, initially introduced in Ref.~\cite{DiVecchia:1981aev}. It takes the following form:
\begin{equation}
    \mathcal{Q}_L = \frac{1}{32\pi^2} \sum_{x} \epsilon_{\mu\nu\rho\sigma} \text{Tr}\left[C_{\mu\nu}(x)C_{\rho\sigma}(x)\right]\,,
    \label{eq:Q_fieldteo_def}
\end{equation}
and uses a discretization of the gauge strength tensor in terms of a "clover leaf" path \(C_{\mu\nu}\), made by the sum of the plaquettes \(P_{\mu\nu}(x)\) centered in \(x\) and with all the possible orientations in the \(\mu\nu\)-plane, i.e.
\begin{equation}
    C_{\mu\nu}(x) = \frac{1}{4}\text{Im}\left[P_{\mu\nu}(x)+P_{\nu,-\mu}(x)+P_{-\mu,-\nu}(x)+P_{-\nu,\mu}(x)\right]\,.
    \label{eq:clover}
\end{equation}
This operator retains its even parity under transformations and displays discretization effects of order \(\mathcal{O}(a^2)\). To smooth the ultraviolet fluctuations of the gauge field and extract the topological charge, we employ the gradient flow method as described in \cite{Luscher:2010iy}. We utilize the standard Wilson action for the smoothing action in the flow equation, with an elementary integration step of \(\epsilon=0.01\). The computation of the topological charge is performed on the smoothed fields at intervals of 
 \(\Delta \tau_{\rm flow}=0.1\). Selecting an appropriate flow time is crucial. It should be large enough to cancel out discretization effects but should also maintain the gauge field's topological properties. Therefore, we investigate the dependence of our final results on \(\tau_{\rm flow}\) in search of a plateau region.  According to Ref.~\cite{Luscher:2010iy}, we anticipate that this plateau emerges when \(a\sqrt{8\tau_{\rm flow}}\sim \mathcal{O}(0.1\text{fm})\). 

In Figure~\ref{fig:topological_charge}, we illustrate the evolution of the topological charge and its distribution across the three $N_f=4$ ensembles. The plots conspicuously do not exhibit signs of topological freezing, strongly suggesting the ergodicity of the simulated Markov-Chain. The topological charge distribution appears to have a Gaussian-like shape, characterized by a relatively narrow spread for the smaller physical volumes of the \texttt{cB4.06.16} and \texttt{cC4.05.24} ensembles, while displaying a much broader distribution for the larger physical volume of \texttt{cB4.06.24}. The presence of a topological charge distribution that closely resembles a Gaussian distribution implies that our computations have effectively probed the topological sectors of the theory. Similar behavior is observed across all other ensembles, both in the context of pure gauge theory and $N_f=2+1+1$.

The gradient flow technique also allows for the establishment of a precise physical scale parameter denoted as $t_0$. The concept of this flow observable was initially introduced in~\cite{Luscher:2009eq,Borsanyi:2012zs}. The determination of the gradient flow time $t_0$ follows this prescription: First, we establish
    \begin{eqnarray}
        F(t) = t^2 \langle E(t) \rangle \, \ {\rm where} \ E(t) = \frac{1}{4} B^2_{\mu \nu} (t)\,,
    \end{eqnarray}
where the field strength $B_{\mu \nu}$ is obtained by evolving $F_{\mu \nu}$ along the direction of flow time. We establish the scale $t_0(c)$ by determining the value of $t$ for which $F(t)|_{t=t_0(c)} = c$.  The choice of $c$ is critical and should satisfy the condition $a \ll \sqrt{8 t_0} \ll L$, where selecting a small $c$ can result in larger lattice artifacts, while opting for a larger $c$ generally leads to increased autocorrelations, as noted in \cite{Bergner:2014ska}. In our case, we opt for $c=0.3$, a commonly used value in lattice QCD calculations.

Figure~\ref{fig:gradient_flow} displays both the evolution of the gradient flow time $t_0/a^2$ and its corresponding distribution across the three $N_f=4$ ensembles. The history plots reveal substantial fluctuations in $t_0$, while the distribution notably approximates a Gaussian shape to a considerable extent.}

\begin{figure}
    \centering
    \includegraphics[height=5.5cm]{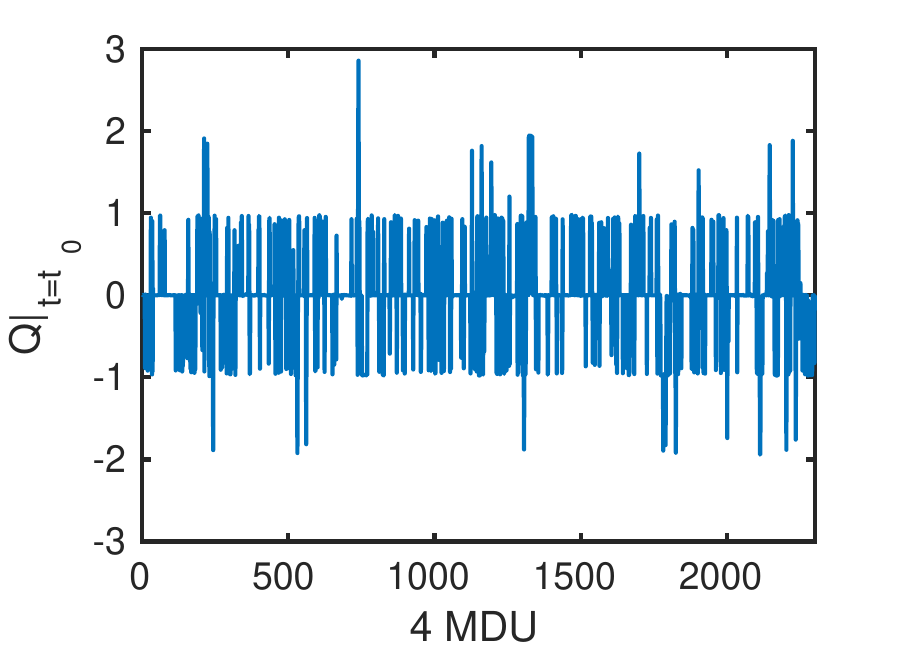}
    \includegraphics[height=5.5cm]{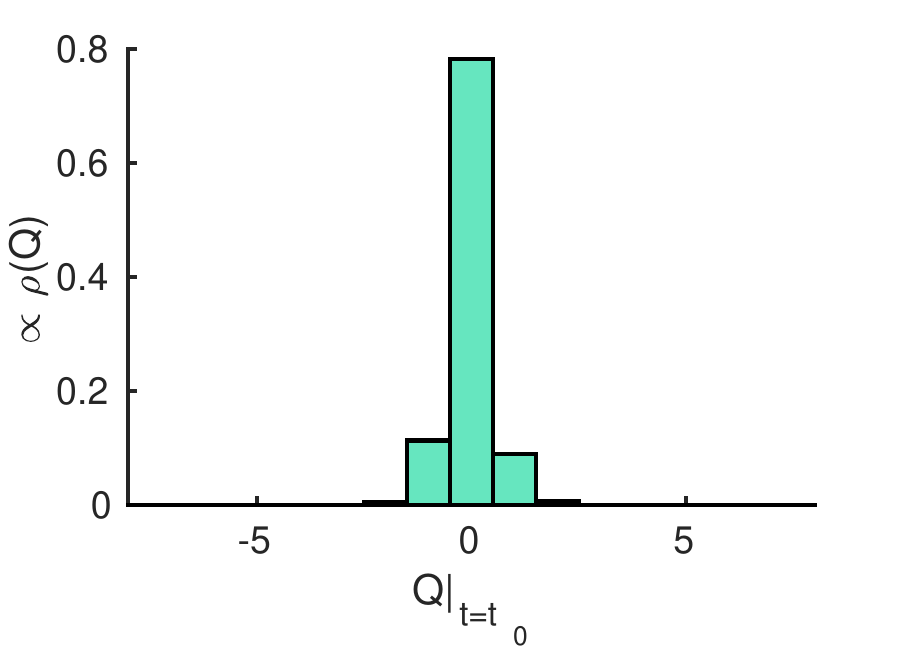}
    \includegraphics[height=5.5cm]{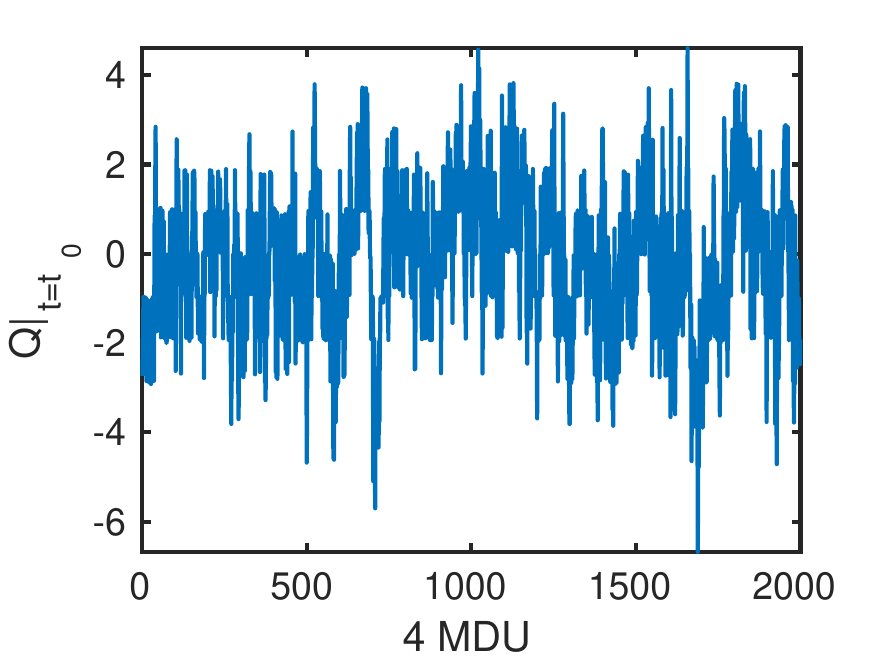}
    \includegraphics[height=5.5cm]{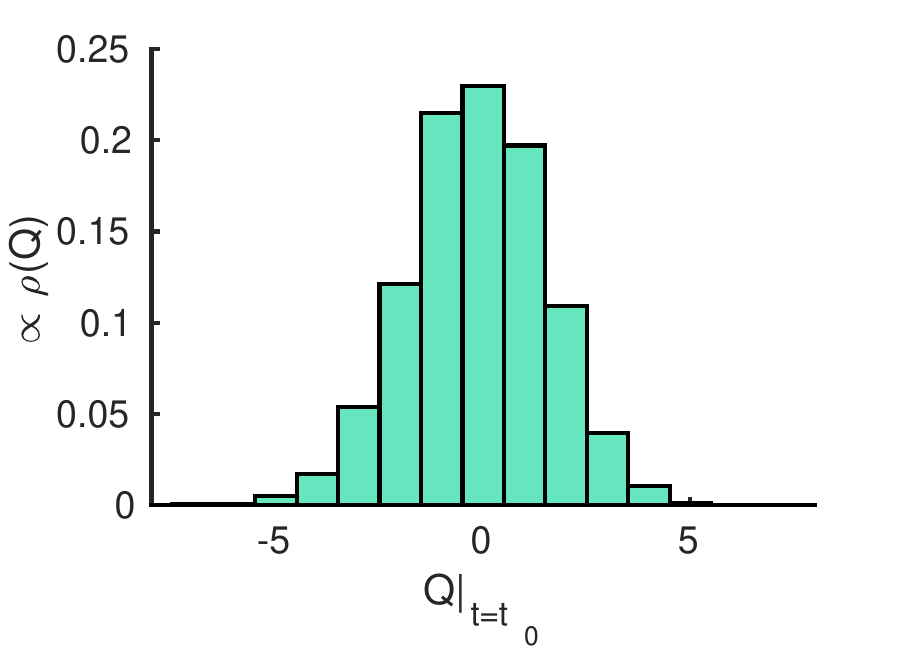}
    \includegraphics[height=5.5cm]{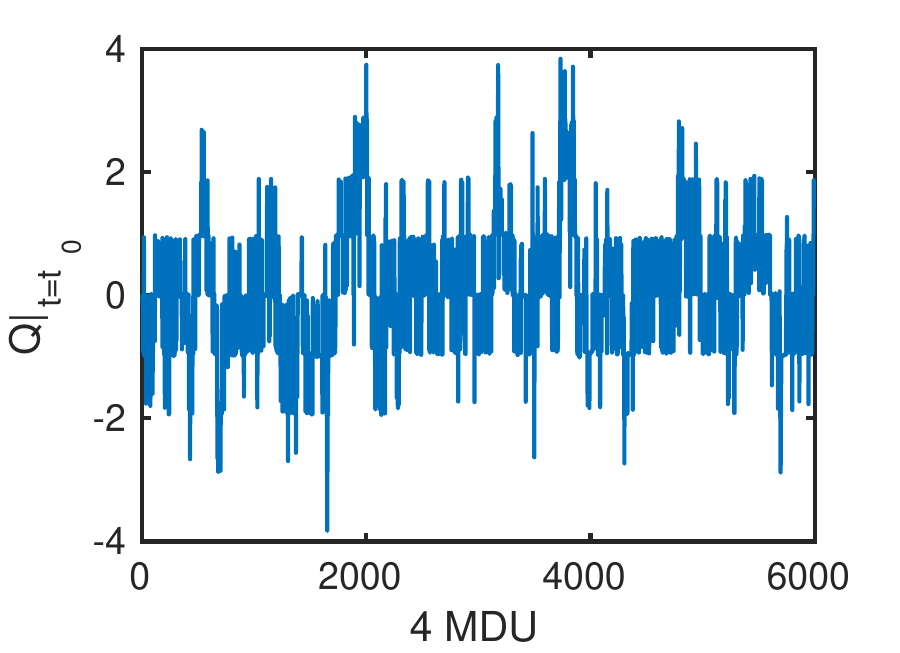}
    \includegraphics[height=5.5cm]{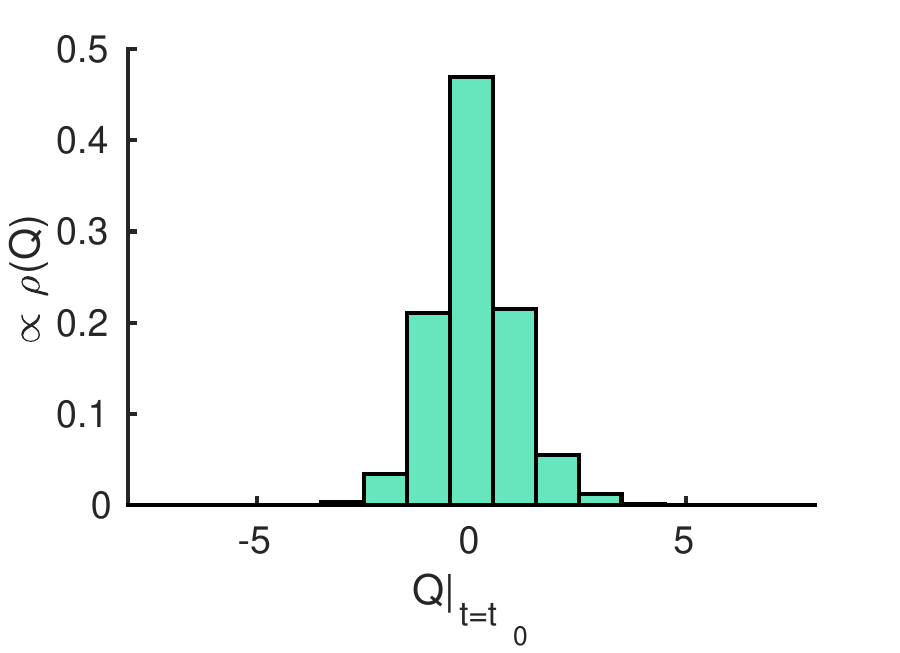}
    \caption{The history of the topological charge as well as its distribution for the three $N_f=4$ ensembles at $t=t_0$. Left column shows the history while right column corresponds to the histogram.}
    \label{fig:topological_charge}
\end{figure}

\begin{figure}
    \centering
    \includegraphics[height=5.5cm]{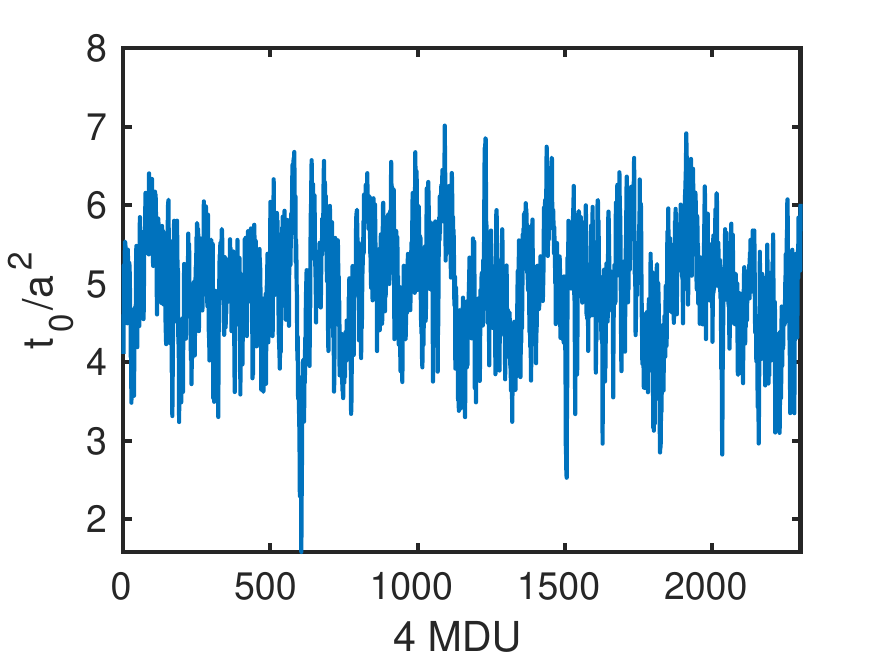}
    \includegraphics[height=5.5cm]{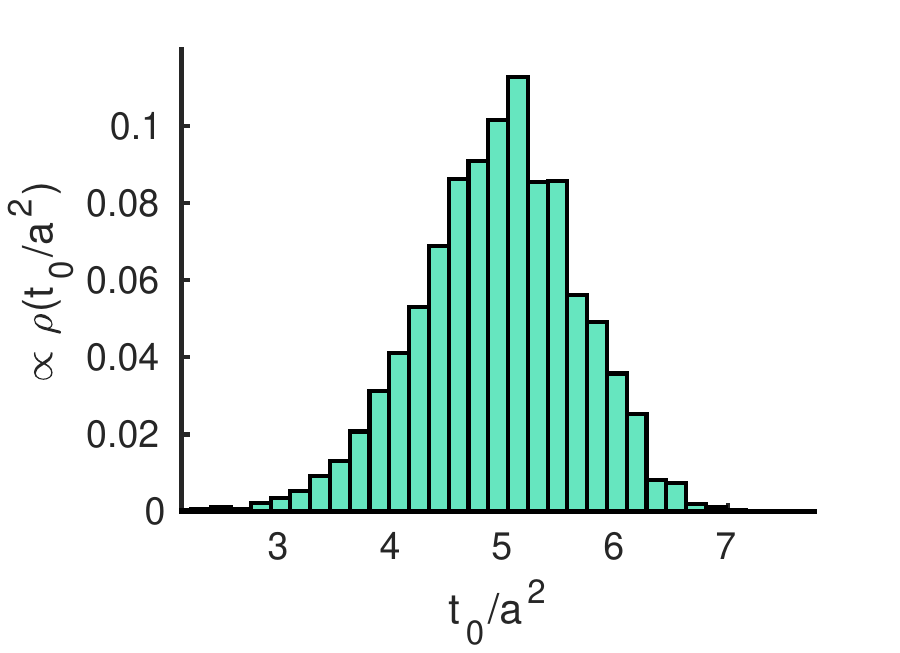}
    \includegraphics[height=5.5cm]{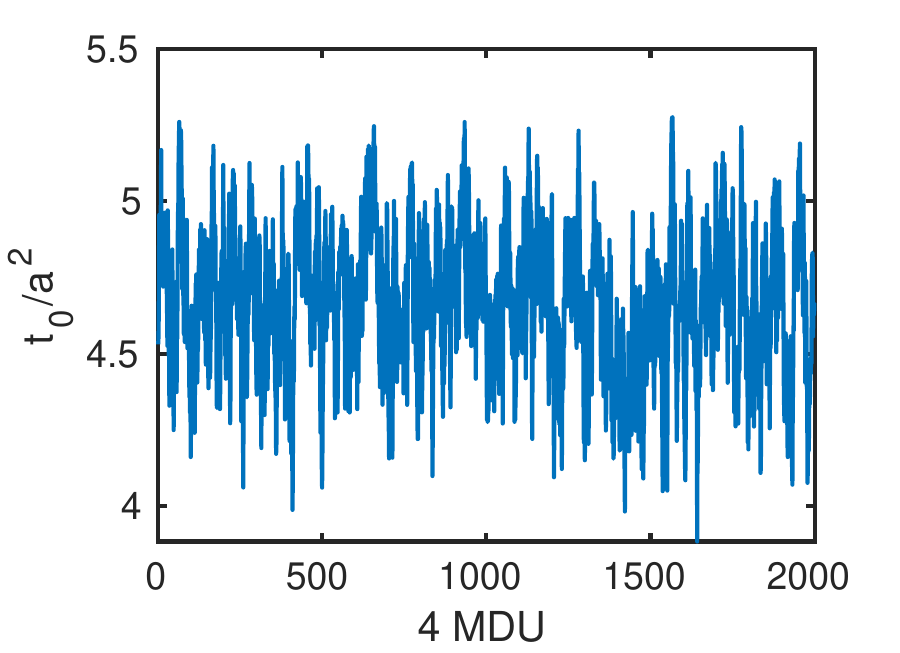}
    \includegraphics[height=5.5cm]{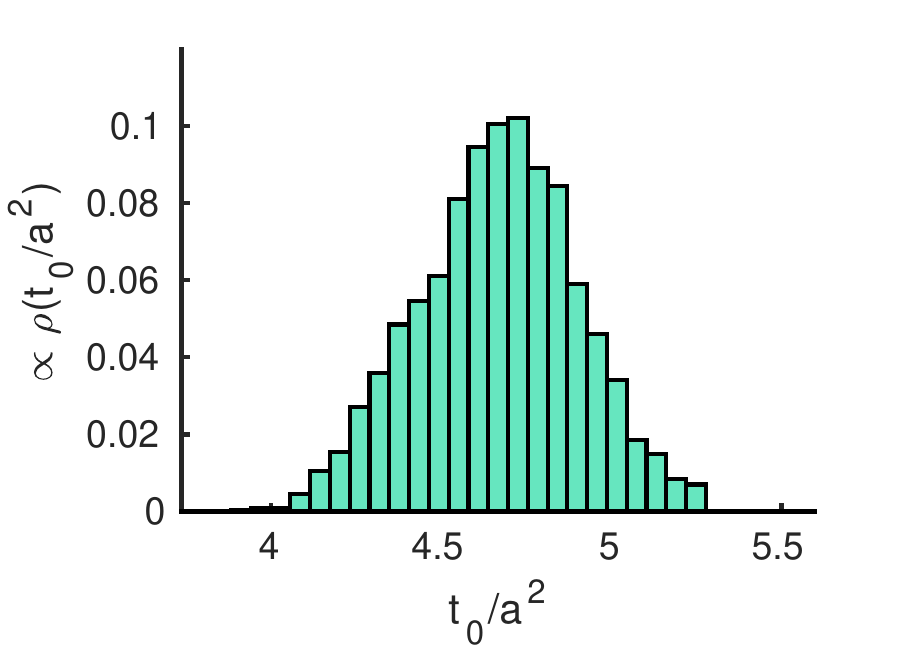}
    \includegraphics[height=5.5cm]{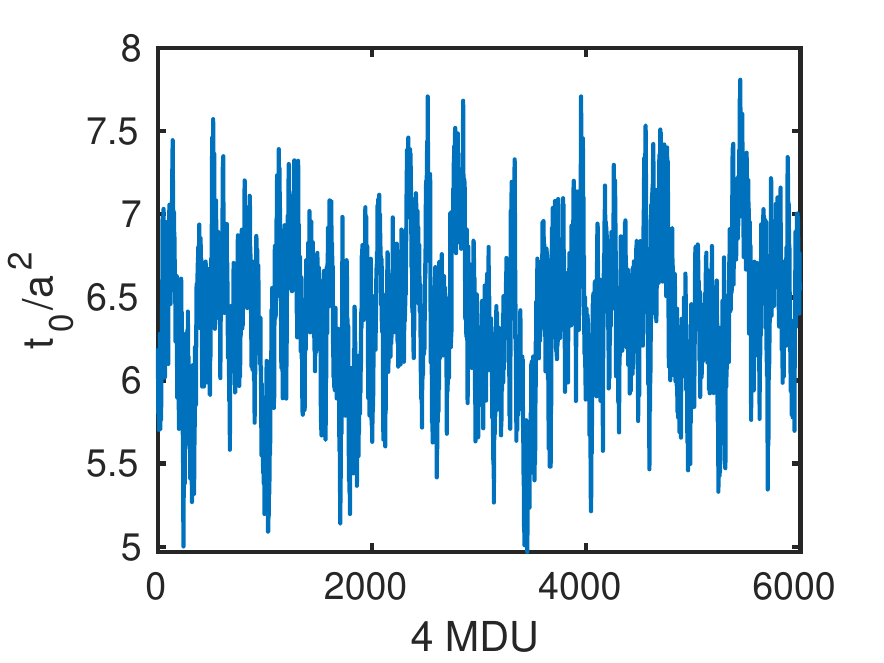}
    \includegraphics[height=5.5cm]{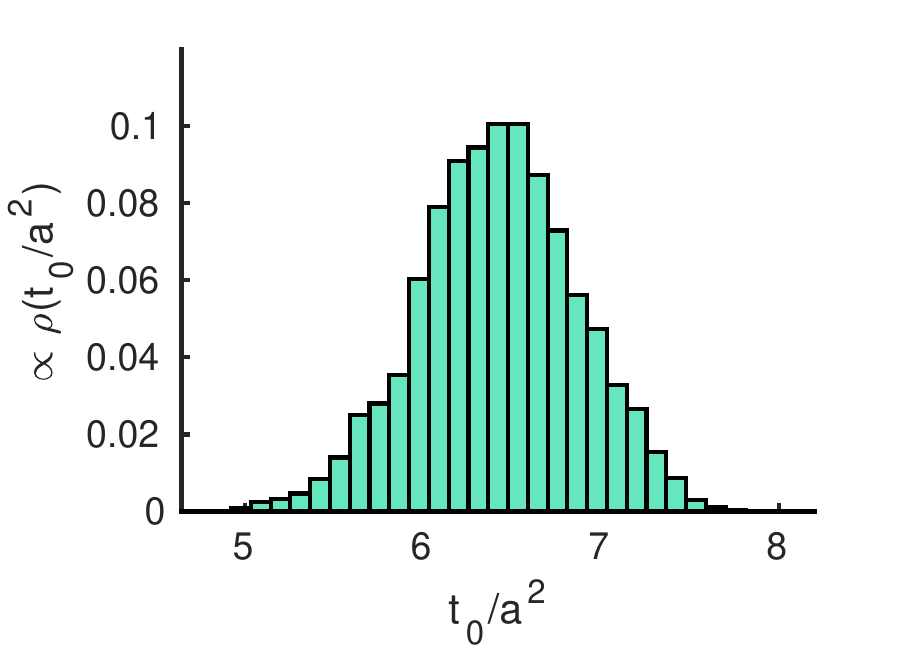}
    \caption{The history of the gradient flow observable $t_0/a^2$
    for the three $N_f=4$ ensembles used for the purposes of this work. On the left we demonstrate the actual measurements while on the right the corresponding histogram.}
    \label{fig:gradient_flow}
\end{figure}

\section{Results}
\label{sec:results}
\subsection{The spectrum of Glueballs in $N_f=4$}
\label{sec:glueballs_Nf4}
We have computed the low-lying spectrum for specific irreducible representations, namely, $A_1^{++}$, $E^{++}$, $T_2^{++}$, and $A_1^{-+}$. These representations correspond to the scalar ($A_1^{++}$), tensor ($E^{++} + T_2^{++}$), and pseudoscalar ($A_1^{-+}$) channels. In Figure~\ref{fig:plots_effective_masses_glueballs}, we present the effective mass plateaus for two of these irreducible representations, namely, $A_1^{++}$ and $T_2^{++}$, specifically for the ensemble labeled as \texttt{cB4.06.24}, which serves as our reference ensemble. One notable observation from the calculation is that the effective mass plateaus set relatively late in time compared to what one would typically observe in pure gauge theory. This, results in an overlap of approximately 30 \% to 50 \%, which is significantly lower compared to the rapid convergence seen in $SU(3)$ pure gauge theory, where overlaps often reach values of around 90 \% to 100 \%. This behavior may be explained by the richness of the Hilbert space in the presence of dynamical quarks, leading to mixings among our operators, including states involving mesons. In an attempt to improve the overlaps with the extracted states, we expanded the variational basis of operators by introducing additional paths, but this led to only marginal improvements.

\begin{figure}[h]
    \centering
    \includegraphics[height=5.4cm]{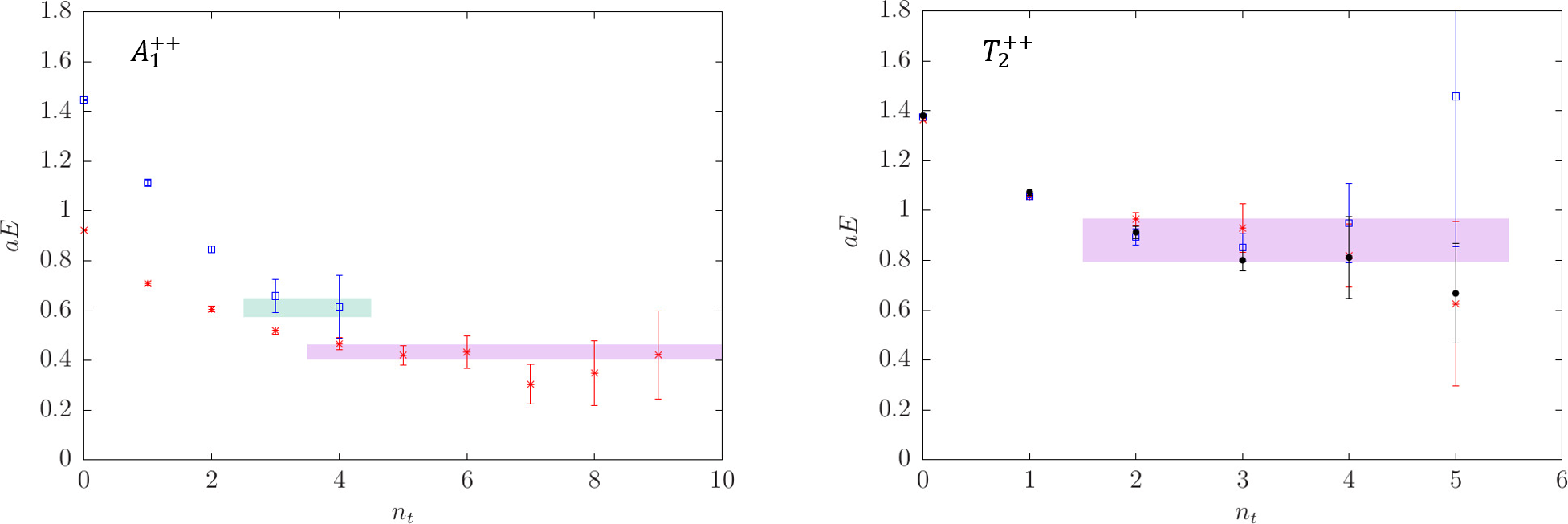} 
    \caption{ \underline{Left panel:} The effective masses corresponding to the ground and first excited states within the irreducible representation $A_1^{++}$ for the ensemble \texttt{cB4.06.24}. \underline{Right panel:} The effective mass corresponding to the ground state of the irreducible representation $T_2^{++}$ for the ensemble \texttt{cB4.06.24}; states in $T_2$ come in triplets, thus, the three states that appear as ground, first excited and second excited states within the variational analysis.}
    \label{fig:plots_effective_masses_glueballs}
\end{figure}

In Figure~\ref{fig:plots_masses_glueballs}, we display the results for glueball masses in $N_f=4$ for three different scenarios: ${\beta=1.778 , L=16}$, ${\beta=1.778 , L=24}$, and ${\beta = 1.836, L = 24}$, presented in the left, middle, and right panels, respectively. Additionally, we provide a comparison with the corresponding values extracted from the pure gauge theory. These plotted data are also tabulated in Tables~\ref{tab:masses_A1++},~\ref{tab:masses_E++},~\ref{tab:masses_T2++}~and~\ref{tab:masses_A1-+} for the representations $A_1^{++}$, $E^{++}$, $T_2^{++}$ and $A_1^{-+}$ respectively. 

The middle panel of Figure~\ref{fig:plots_masses_glueballs}, or Figure~\ref{fig:spectrum_L24_Beta1.778}, serves as our reference plot because we know that finite volume effects do not significantly impact the pion mass for this specific ensemble. Nevertheless, our findings indicate that glueball masses appear to be relatively insensitive to finite volume effects. This observation becomes evident when comparing the middle panel, where the spatial lattice size is $L=24$, with the left panel, where $L=16$, while keeping the value of $\beta$ constant. However, it's worth noting that reducing the volume can introduce other finite volume effects, such as non-glueball states known as di-torelons.

Indeed, in the left panel of Figure~\ref{fig:plots_masses_glueballs}, it is apparent that for the representation $E^{++}$, the ground state seems to be a di-torelon with a mass slightly greater than twice that of the torelon, as corroborated by the data in Tables~\ref{tab:masses_E++} and~\ref{tab:masses_string}. It's important to note that this is a somewhat heuristic way of identifying the nature of this state as a di-torelon. A more rigorous approach to confirming whether this state is indeed a di-torelon involves the following procedure. Initially, we expand our variational basis to include operators deliberately designed to maximize their overlap with di-torelon states. These operators are constructed by taking the product of a spatial Polyakov loop with its charge conjugate, ensuring that the two lines have opposite directions. We then form linear combinations of these spatial double lines to encode the correct quantum numbers for the $E^{++}$ representation. At first, we conduct the analysis using the complete basis of operators, which includes the di-torelon operators. Surprisingly, the resulting spectrum shows no discrepancies when compared to the spectrum obtained using the basis that excludes the di-torelon operators. Subsequently, we carry out a similar analysis using only the di-torelon operators. In this case, the spectrum obtained confirms the existence of the suspected di-torelon states, but intriguingly, the lightest state, typically identified as a glueball, does not appear. As anticipated, as we increase the lattice's spatial length from $L=16$ to $L=24$, the di-torelon state vanishes.

Furthermore, we conducted an examination to determine whether decreasing the lattice spacing would have an impact on the spectrum of $N_f=4$. This investigation demonstrated that, particularly for the lower-energy spectrum, the spectrum depicted in the middle plot of Figure~\ref{fig:plots_masses_glueballs} accurately represents the continuum physics. This assertion becomes evident when comparing the spectrum in the middle panel to that in the right panel of Figure~\ref{fig:plots_masses_glueballs}, where we vary the value of $\beta$.

\begin{figure}[h]
\vspace{-0.5cm}
    \centering
   \scalebox{1.25}{\hspace{-0.80cm}
    \includegraphics[height=7.5cm]{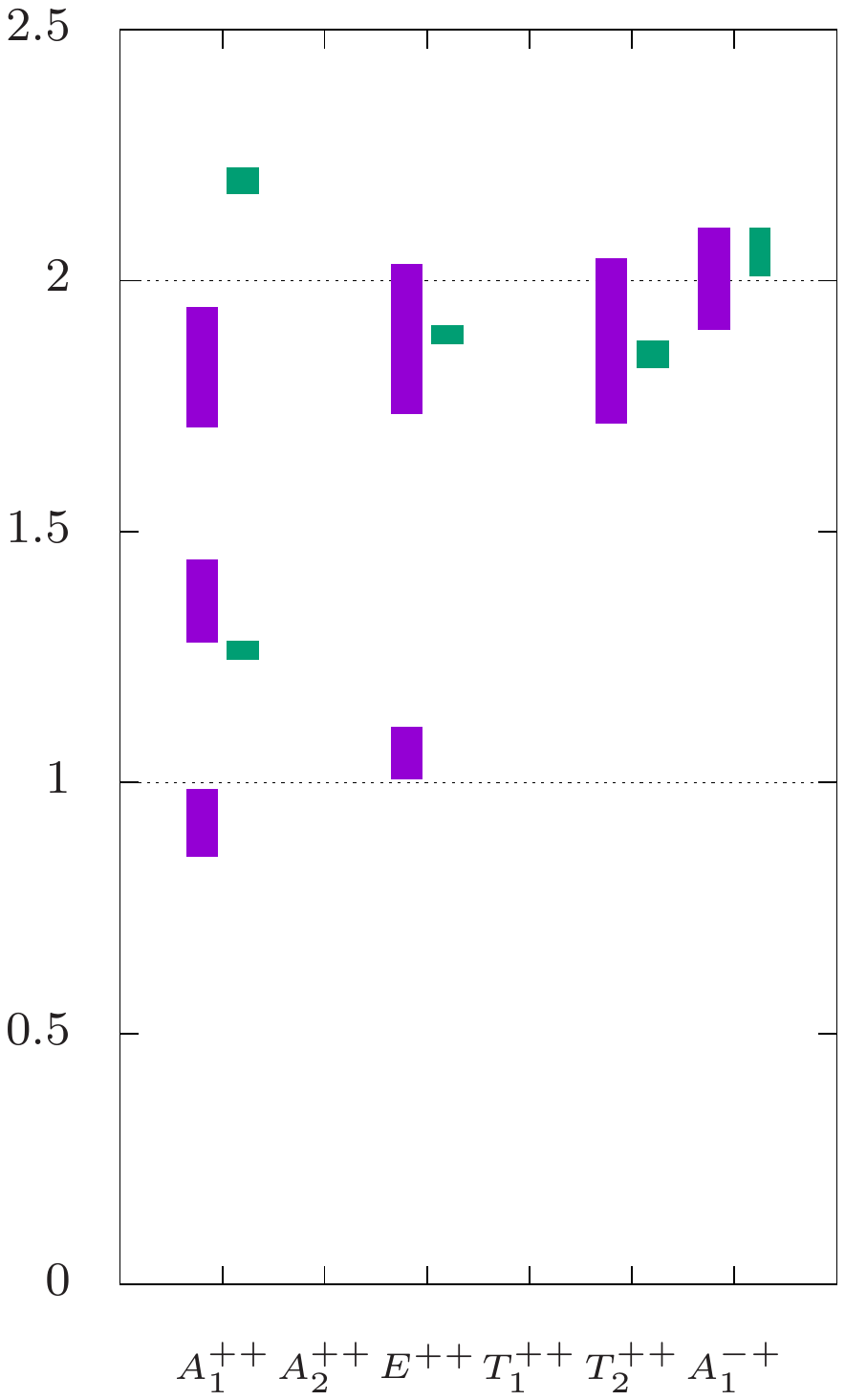} \hspace{-1.4cm}\includegraphics[height=7.5cm]{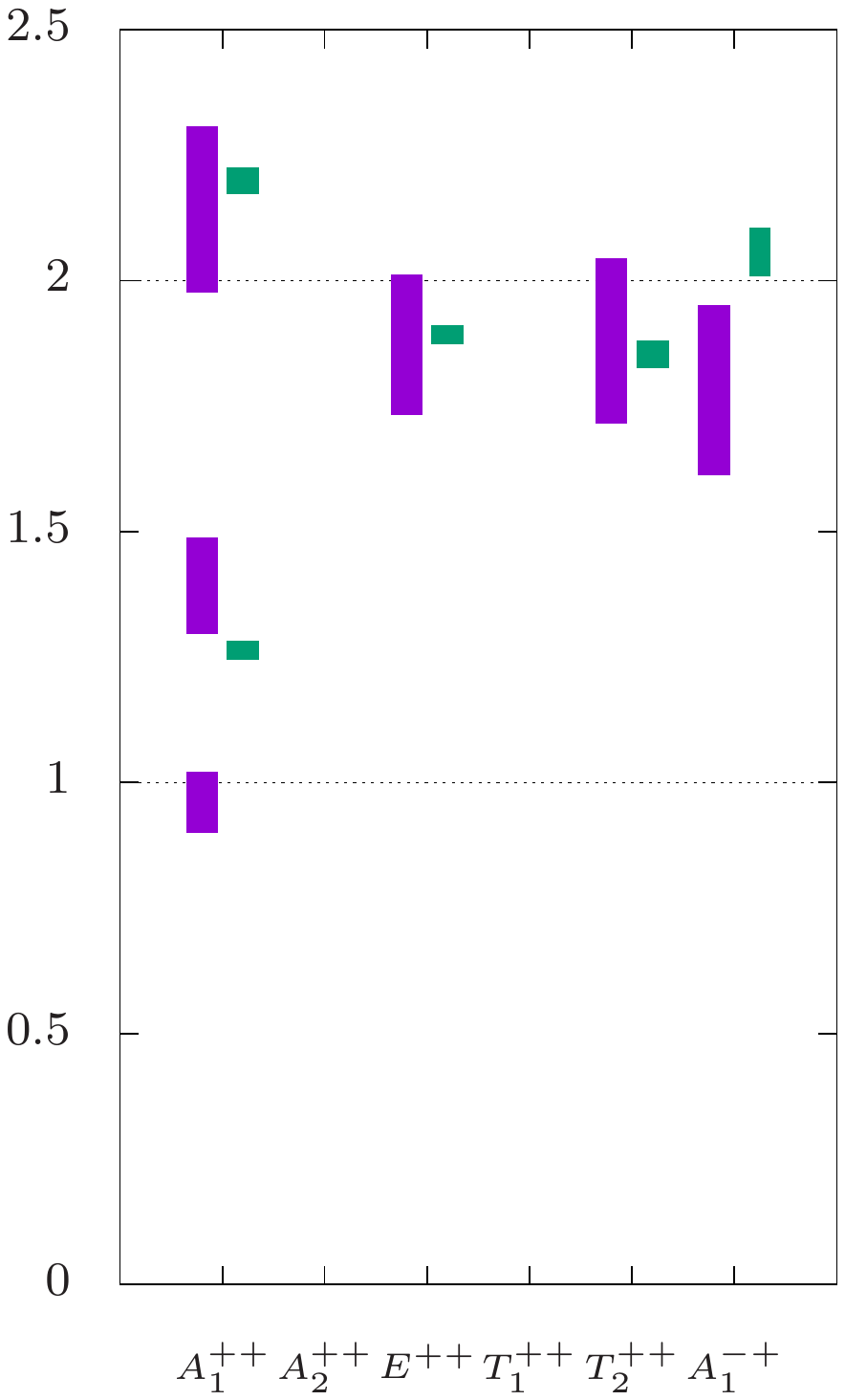}
    \hspace{-1.4cm}\includegraphics[height=7.5cm]{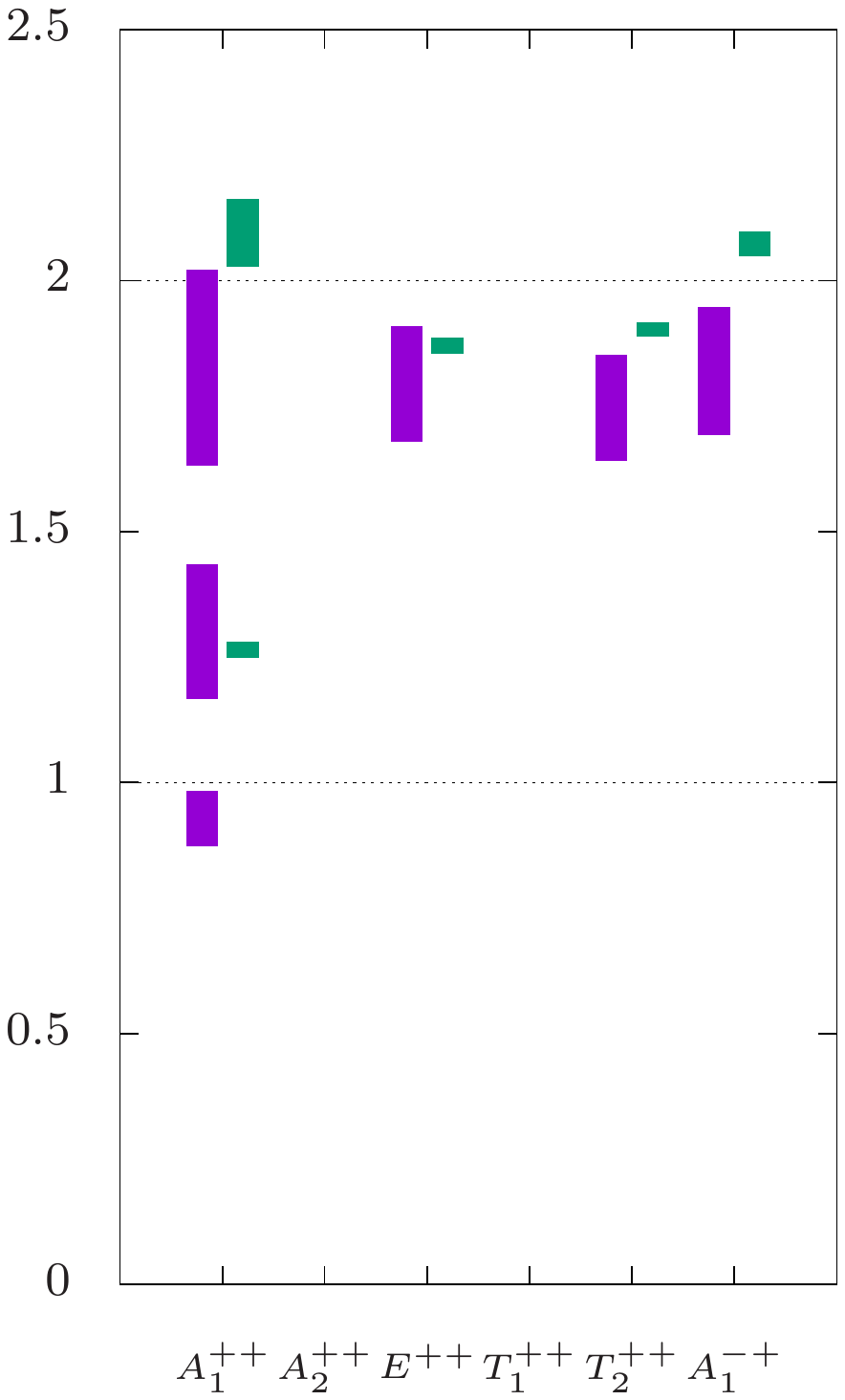} \put(-405,140){\scriptsize $M\sqrt{t_0}$}}
    \caption{The spectrum of glueballs for the representations $A_1^{++}$, $E^{++}$, $T_2^{++}$, $A_1^{-+}$ for the ensembles \texttt{cB4.06.16}(left panel), \texttt{cB4.06.24}(middle panel) 
    and \texttt{cC4.05.24}(right panel). In purple we denote the states of $N_f=4$ QCD while in green the states of $SU(3)$ pure gauge.}
    \label{fig:plots_masses_glueballs}
\end{figure}
Let's now focus to the comparison between $N_f=4$ QCD and pure $SU(3)$ gauge theory. To achieve this, we begin by determining the value of $t_0/a^2$ for the $N_f=4$ configurations. Following this, we proceed to extract the glueball spectrum for the pure gauge theory at a specific $\beta$ value that corresponds to the same $t_0/a^2$. You can find the values of $t_0/a^2$ in Table~\ref{tab:params_sim_Nf2+1+1}, while the corresponding $\beta$ values have been determined through interpolations of data from references such as Refs~\cite{Luscher:2010iy,Francis:2015lha,Luscher:2011kk,Ce:2015qha}, and they are presented in Table~\ref{tab:params_sim_pure_gauge} and graphically represented in Figure~\ref{fig:t_0_beta_interpolation}.

In Figure~\ref{fig:plots_masses_glueballs}, the glueball masses for pure $SU(3)$ gauge theory are depicted in green, while those for $N_f=4$ are represented in purple. Notably, a distinguishing feature is the presence of an additional state within the spectrum of states for the $A_1^{++}$ channel. Specifically, the ground state of the $SU(3)$ gauge theory in the $A_1^{++}$ channel seems to differ from that of $N_f=4$ and instead aligns more closely with the first excited state in the $N_f=4$, $A_1^{++}$ channel. This observation immediately raises the hypothesis that the ground state in $N_f=4$ incorporates some quark content, while the first excited state corresponds to a glueball state. However, it's important to note that this hypothesis requires further investigation.

Shifting our attention to the ground states of $E^{++}$ and $T_2^{++}$, which correspond to the ground state of $J^{PC}=2^{++}$, it's noteworthy that there is a consistent agreement between $N_f=4$ and pure gauge theory. This suggests that conducting precise investigations of the tensor glueball in the pure gauge context can serve as a reliable benchmark for comparisons with experimental data. Lastly, similar trends have emerged for the pseudoscalar ground state $A_1^{-+}$. In particular, it appears to exhibit a slightly lower mass when dynamical quarks are included in the simulation, with a deviation of one sigma. Additionally, the ground state of $2^{++}$ in the $N_f=4$ context closely resembles that of $A_1^{-+}$, mirroring the situation observed in pure gauge theory.
\begin{figure}[h]
\vspace{-0.5cm}
    \centering
    \scalebox{1}{\includegraphics[height=8cm]{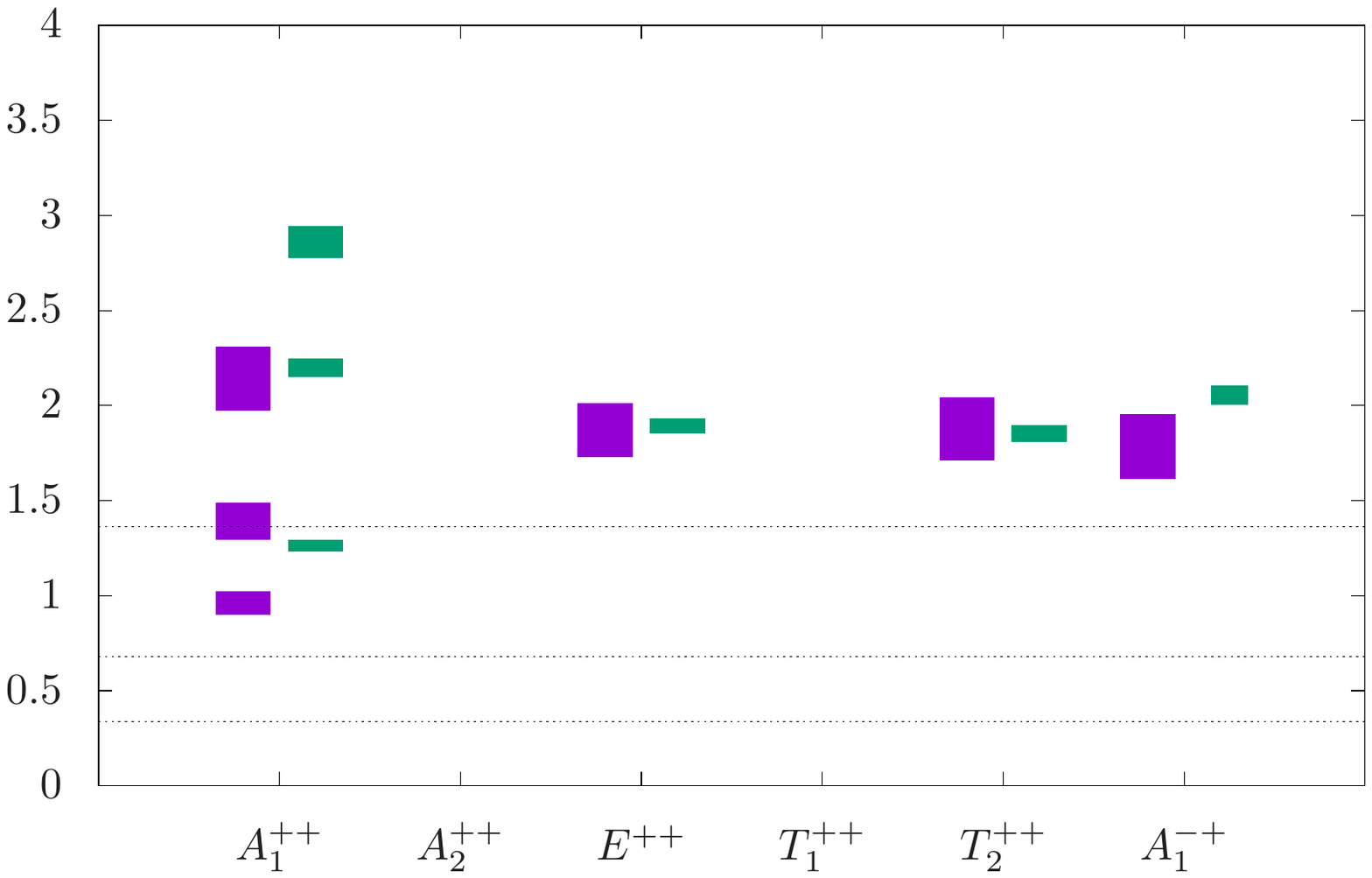}\put(-320,145){$M\sqrt{t_0}$}\put(0,40){$ m_{\pi}\sqrt{t_0}$}\put(0,57){$ 2 \times m_{\pi}\sqrt{t_0}$}\put(0,80){$ 4 \times m_{\pi}\sqrt{t_0}$}}
    \caption{The spectrum of glueballs for the representations $A_1^{++}$, $E^{++}$, $T_2^{++}$, $A_1^{-+}$. In purple we denote the states of $N_f=4$ QCD for the ensemble \texttt{cB4.06.24} while in green the states of $SU(3)$ pure gauge for $\beta=6.117$. The dashed lines correspond to 1, 2 and 4 times the pion mass from bottom to top respectively.}
    \label{fig:spectrum_L24_Beta1.778}
\end{figure}

\section{On the content of the $A_1^{++}$ ground state}
\label{sec:Nf211}
One of the most intriguing aspects of the obtained spectra is the appearance of an additional state in the scalar channel, as indicated in Figure~\ref{fig:plots_masses_glueballs}. To gain a deeper understanding of the nature of this extra state within the $A_1^{++}$ irreducible channel, we conducted a similar investigation for $N_f=2+1+1$ configurations. Specifically, we extracted the low-lying spectrum for $A_1^{++}$ under the more physically relevant conditions of $N_f=2+1+1$ twisted mass fermions, considering two distinct values of pion masses, approximately $260$ and $350$ MeV. Once again, our analysis involved ensembles composed of a substantial number of configurations, although not as numerous as those for $N_f=4$. For this investigation, we deemed the ensembles \texttt{cA211.53.24} (comprising 5000 configurations) and \texttt{cA211.25.32} (consisting of 2500 configurations) to be suitable choices. You can find detailed information about these two ensembles in Table~\ref{tab:params_sim_Nf2+1+1}.
\begin{figure}[h]
\vspace{-0.0cm}
    \centering
    \includegraphics[height=8cm]{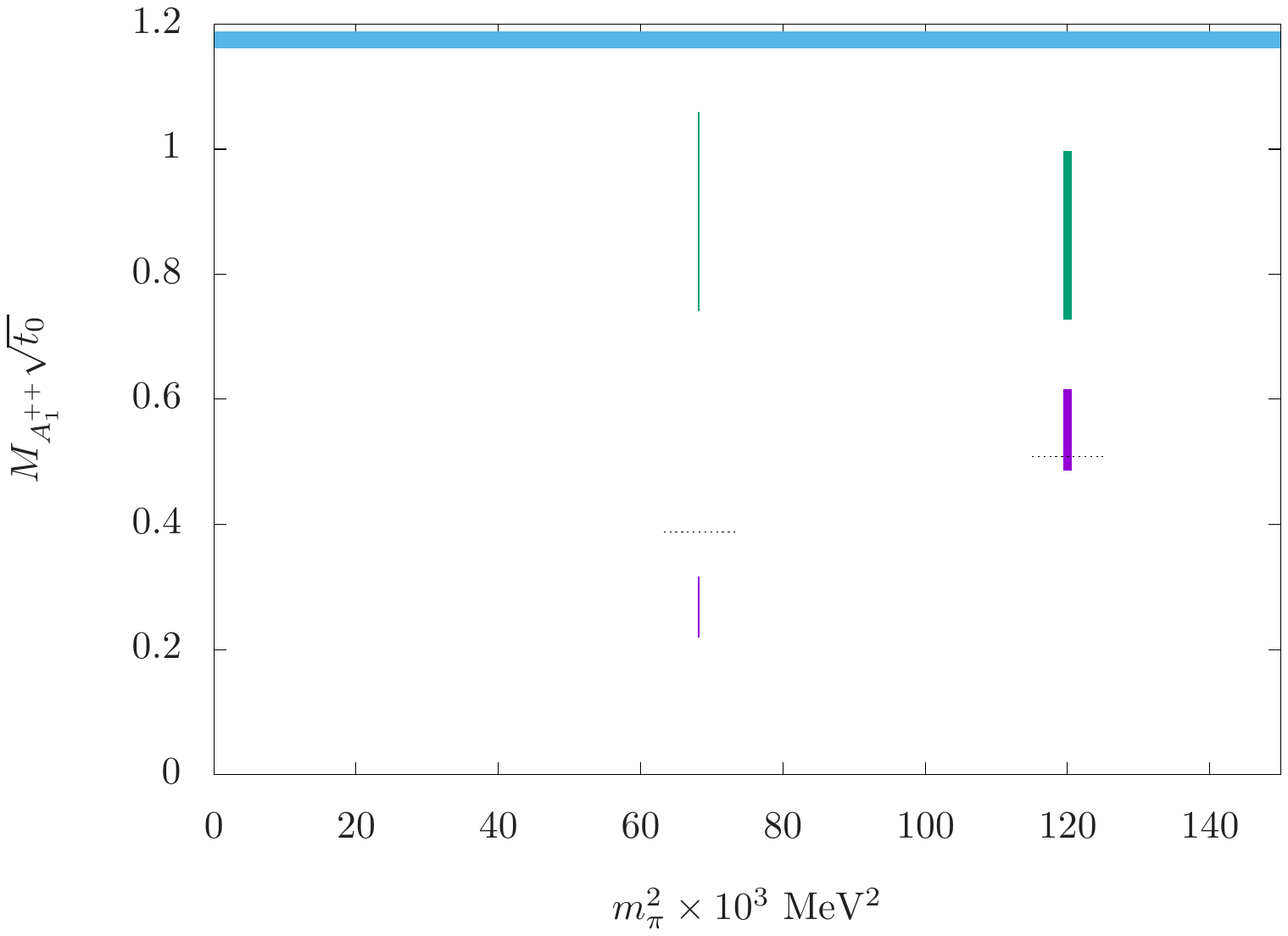}
    \caption{The ground state and the first excited state of the scalar channel $A_1^{++}$ for the ensembles \texttt{cA211.53.24}  and \texttt{cA211.25.32}. The blue band represents the scalar glueball mass of the $SU(3)$ pure gauge theory at $\beta=5.917$ $(t_0/a^2 \sim 2.38)$ extracted for the purposes of this work. The black dashed lines correspond to twice the pion mass.}
    \label{fig:spectrum_Nf211}
\end{figure}

Figure~\ref{fig:spectrum_Nf211} illustrates the outcomes for both the ground and the first excited states in the $A_1^{++}$ scalar channel for the $N_f=2+1+1$ configurations. In this visualization, the blue band represents the scalar glueball mass in pure $SU(3)$ gauge theory at $\beta=5.917$ ($t_0/a^2 \approx 2.38$), which has been determined specifically for this study. Meanwhile, the dashed black lines indicate approximately twice the mass of the pion for each distinct ensemble.

The message conveyed by this plot is quite evident: the behavior of the ground state strongly correlates with the pion mass, while the first excited state remains relatively consistent. This strongly suggests that the ground state contains quark content, possibly signifying a decay process involving a glueball decaying into either two or four pions. Notably, the mass of the ground state closely aligns with twice the mass of the pion (as indicated by the dashed lines). In contrast, the first excited state, which appears to remain stable, is indicative of representing the glueball mass. It's important to acknowledge that the coarseness of the lattice introduces substantial influences from lattice artifacts, leading to a considerable deviation from the pure gauge theory value. 

\section{The Torelon Ground State and the string tension}
\label{sec:torelon}
We have extracted the ground state of the spectrum of the torelon for the three $N_f=4$ ensembles. The torelons are short enough, enabling us to extract adequate results with good statistics. The effective mass plots resulting from the torelon correlators are shown in Figure~\ref{fig:plots_masses_torelons}. As in the case of glueballs, it is stiking that the mass plateau sets late in time, resulting to a low overlap of $20 \%$; nevertheless the plateaus are quite convincing, leading to a precise estimation of the torelon masses. The closed flux-tube masses are presented in the second column of Tab.~\ref{tab:masses_string}. Assuming that the torelon behaves approximately as a bosonic string with energy described by Eq.~\ref{eq:nambu_goto}, we extracted the associated string tension which is presented in the third column of Tab.~\ref{tab:masses_string}. The string tension for $L=16$ and $L=24$ and $\beta=1.778$ is compatible within statistics suggesting that Eq.~\ref{eq:nambu_goto} provides a good approximation for torelon energies for the intermediate values of lengths between $L=16$ and $L=24$. In Fig.~\ref{fig:plot_nambu_goto} we provide the prediction of Eq.~\ref{eq:nambu_goto} with string tension fixed to the value extracted for $L=24$; both points lay on the Nambu-Goto prediction.
\begin{figure}[h]
    \centering
    \scalebox{1.13}{\hspace{-1cm}
    \includegraphics[height=4.8cm]{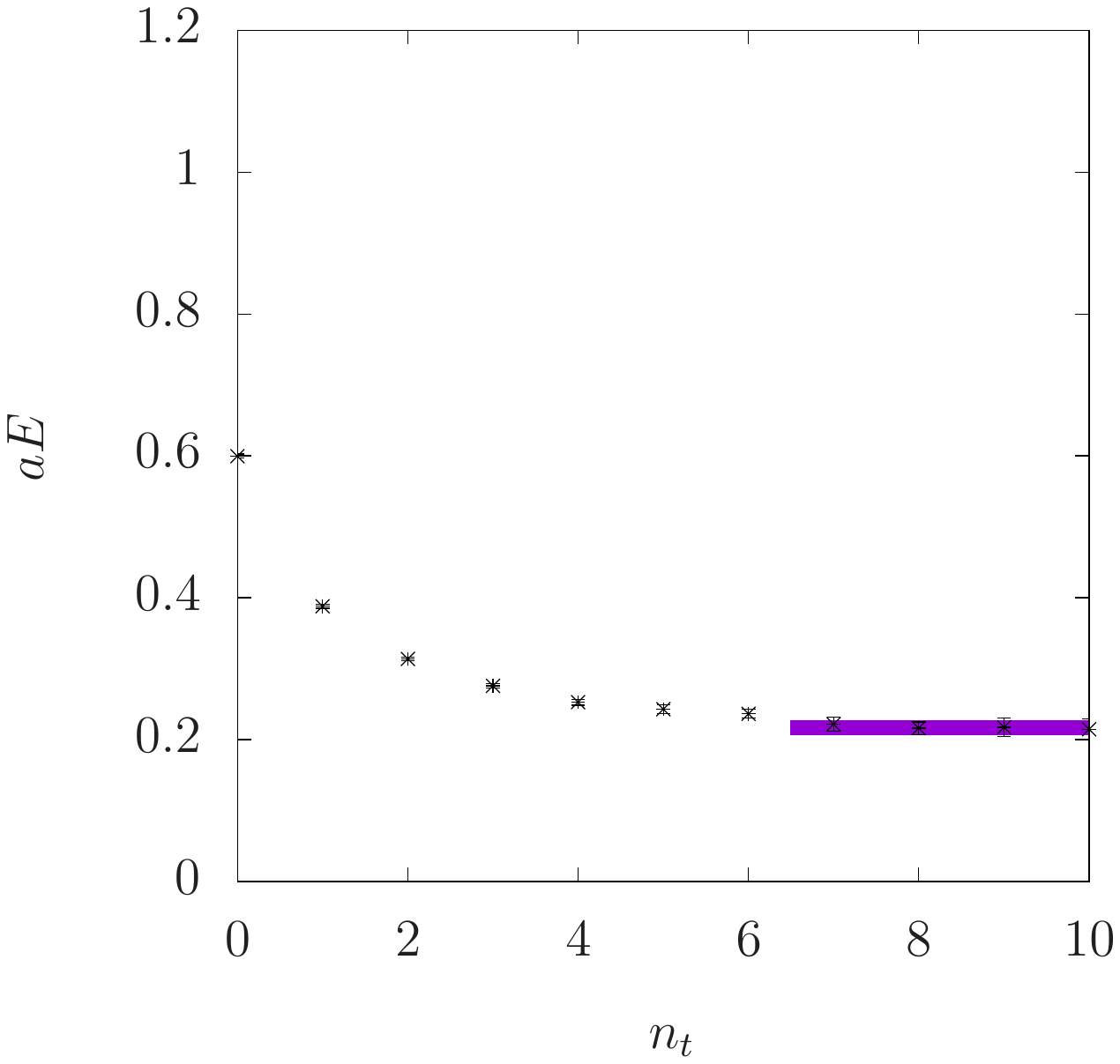} \hspace{-1.5cm}\includegraphics[height=4.8cm]{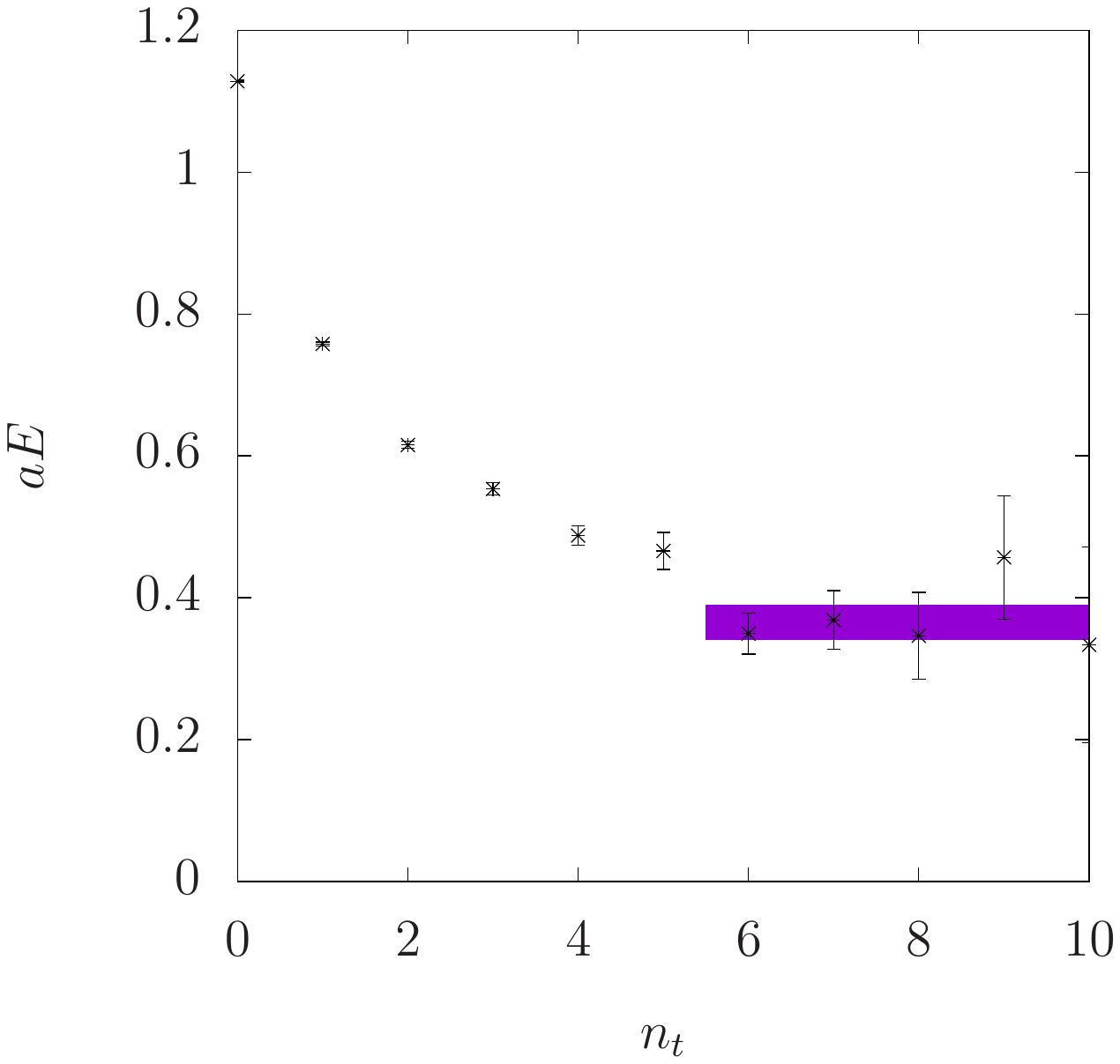}
    \hspace{-1.5cm}\includegraphics[height=4.8cm]{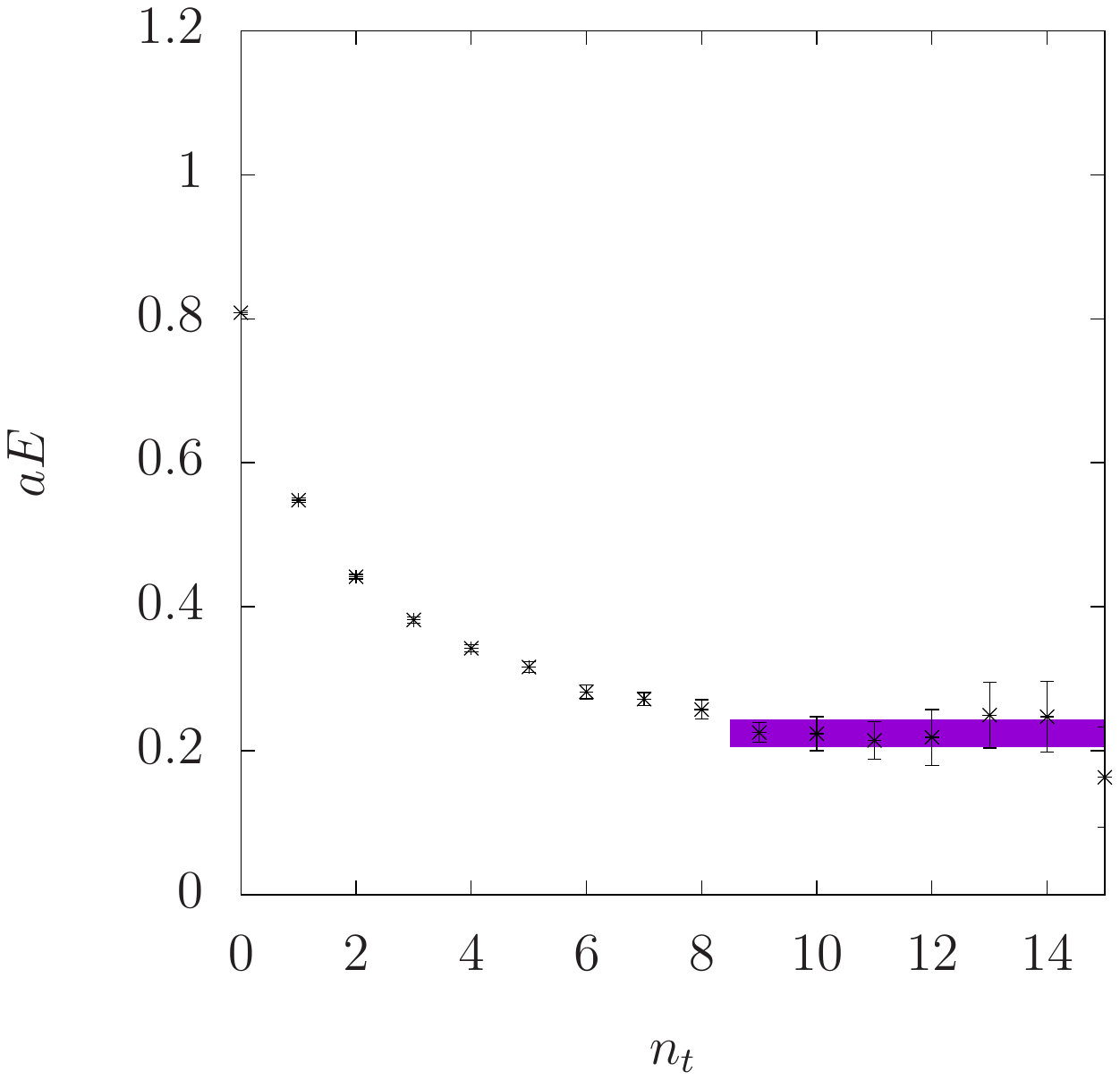}}
    \caption{The effective mass of the torelon ground state for the ensembles \texttt{cB4.06.16}(left panel), \texttt{cB4.06.24}(middle panel) 
    and \texttt{cC4.05.24}(right panel).}
    \label{fig:plots_masses_torelons}
\end{figure}
A careful look at the masses of the torelon as well as the pion masses shown on the sixth column of Tab. \ref{tab:params_sim_Nf4}, indicates that for \texttt{cB4.06.24}, the torelon ground state is larger than twice the pion mass, and in accordance with Nambu-Goto. Furthemore, the first excited state of torelon for the same ensemble is much larger $\sim 0.9508(870)$.  This disfavors the scenario of having a complete string breaking of the flux-tube to two pions. Of course possible string breaking phenomena could have an effect on the dynamics of the flux-tube and drive the string tension to lower values. As a matter of fact, if we compare the string tension extracted for $N_f=4$
QCD compared to the string tension extracted in pure gauge theory for the same values of $t_0/a^2$ (Tab.~\ref{tab:params_sim_pure_gauge}) we observe that the dynamical fermions induce a suppression in $a^2 \sigma$ of 50-70 \%. At this point we come across the subtlety on what we should be using as a valid scale in order to give physical units in the quantities being measured. The two options we can consider are the Wilson flow time and the string tension; this question requires dedicated investigation. However, for the purposes of this work it looks more suitable to use the Wilson flow time which appears to result to gluonic quantities such as glueball masses affected only negligibly due to the inclusion of dynamical quarks.    
\begin{figure}[h]
    \centering
    \scalebox{1.0}{\includegraphics[height=4.8cm]{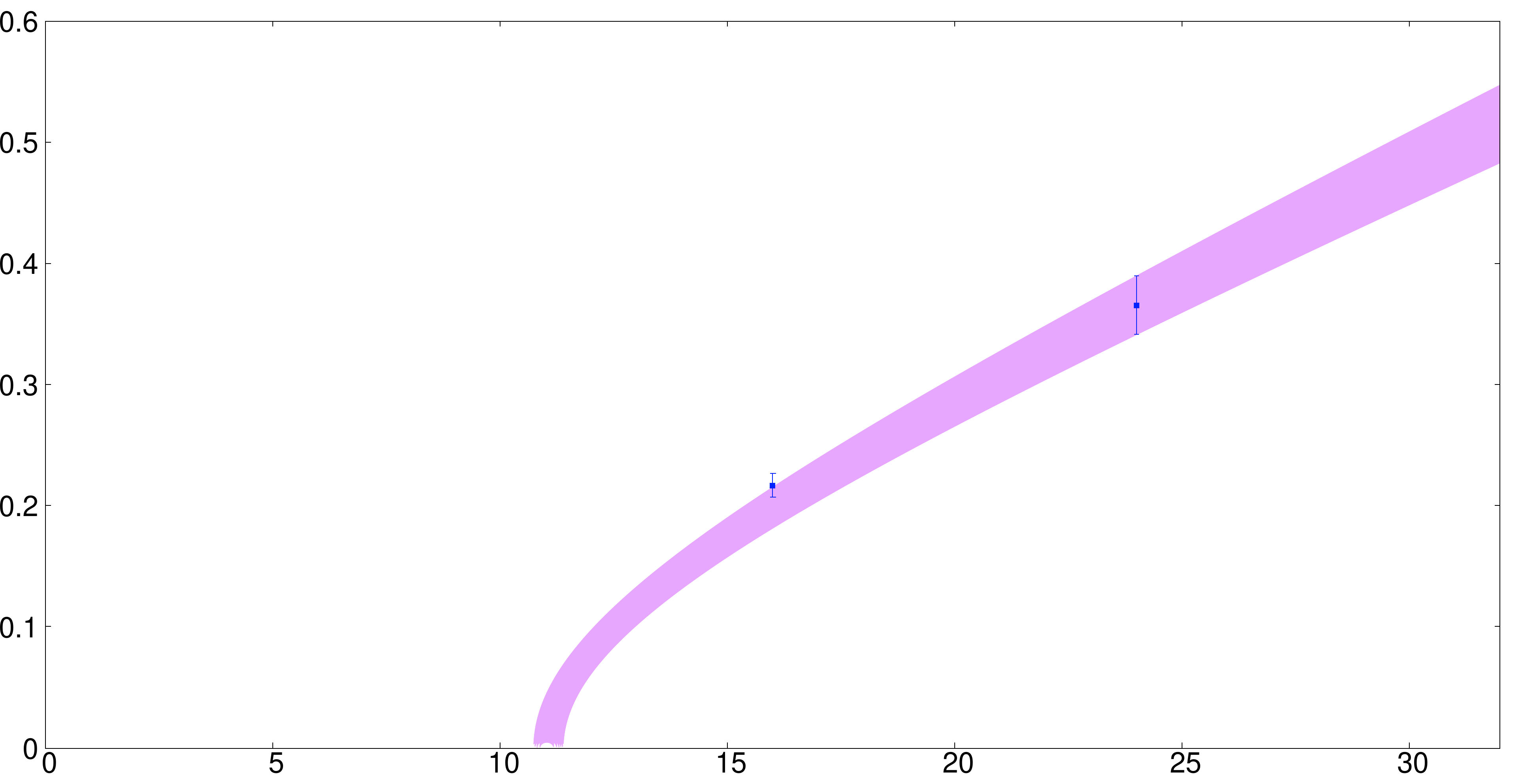}\put(-140,-10){$l$}\put(-285,70){$aE$}}
    \caption{The energy of the torelon ground state for $l=16a$ and $l=24a$ for $N_f=4$ and $\beta=1.778$ as well as the Nambu-Goto formula of Eq.~\ref{eq:nambu_goto} with string tension taken from the estimation extracted for $L=24a$.}
    \label{fig:plot_nambu_goto}
\end{figure}
Focusing now on the ensemble \texttt{cB4.06.16}, we observe that the value of the torelon mass is $am_T \sim 0.22$ while the potential di-torelon state is $am_{DT} \sim 0.51$ (Tab.~\ref{tab:masses_E++}). This is in accordance with the scenario of the ground state of $E^{++}$ representation corresponding to the di-torelon, given that it is slightly larger than twice the mass of the torelon.

\section{Comparison with other results}
\label{sec:comparison_with_other_results}

In this section we provide a comparison of our results for the scalar, tensor as well as pseudoscalar glueball masses with other results from the literature. In this comparison we do not consider the $0^{++}$ additional lightest state which appears to have quark content. 
 
We begin with early studies on the unquenched glueball spectrum. In 2001 the authors of Ref.~\cite{McNeile:2000xx} used for first time, in addition to pure gluonic operators, also $q {\bar q}$ operators which were included in the variational analysis and the mass of the flavor-singlet state was measured in a system with $N_f=2$ flavors of clover improved Wilson fermions. The results obtained out of this work demonstrated a suppression on the flavor-singlet scalar mass compared to our work. 

\begin{figure}[h]
    \centering
    \scalebox{1.0}{\includegraphics[height=9.0cm]{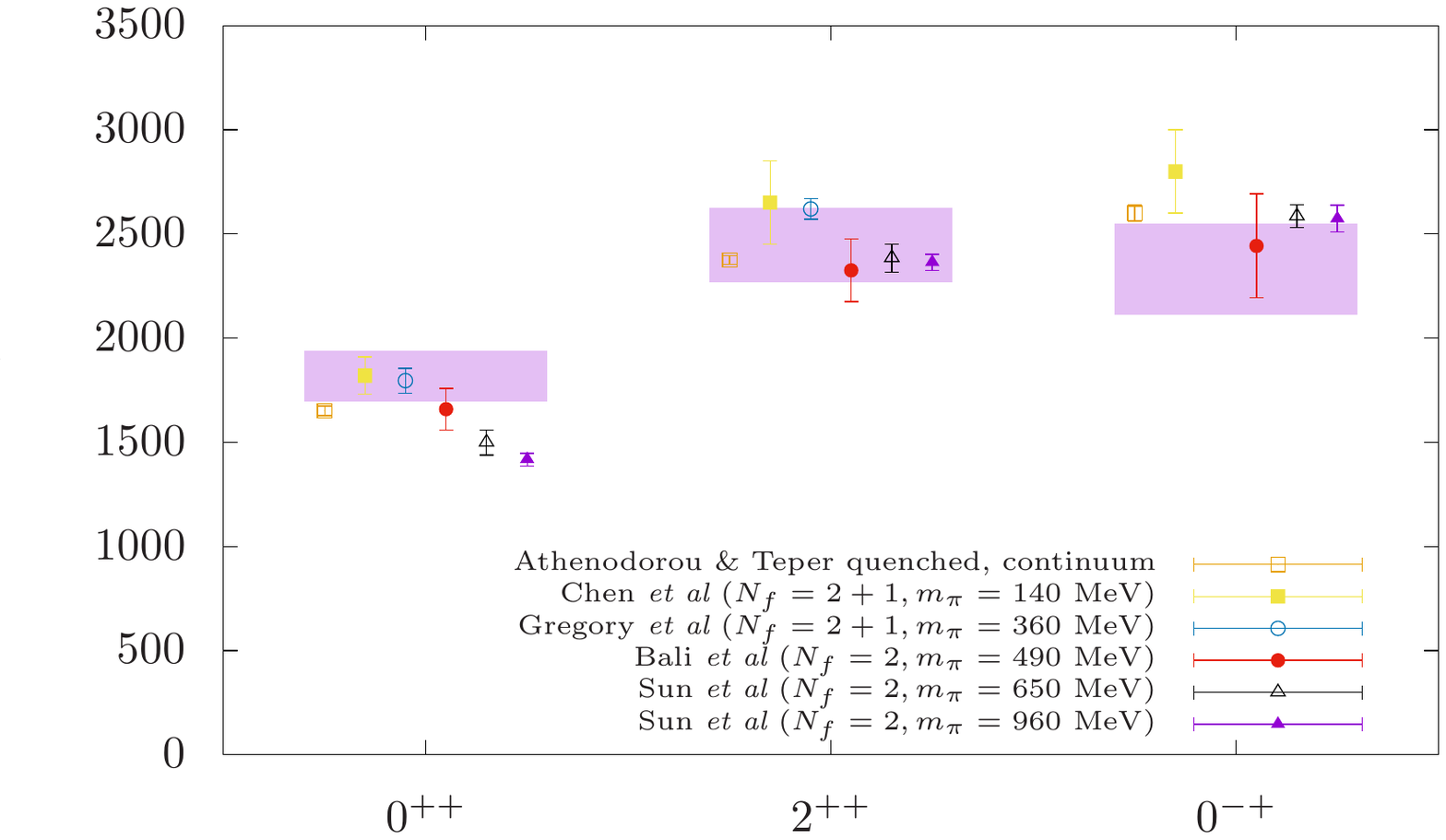}\put(-143,10){$J^{PC}$}\put(-335,150){MeV}}
    \caption{Comparison of our results for the ground states of scalar ($0^{++}$), tensor ($2^{++}$) and pseudoscalar ($0^{-+}$) denoted by the magenta bands with other unquenched results~\cite{Chen:2022goa,Gregory:2012hu,Bali:2000vr,Sun:2017ipk} as well as the quenched results from~\cite{Athenodorou:2020ani}.}
    \label{fig:plot_comparison_unquenched}
\end{figure}

In Ref. \cite{Gregory:2012hu}, the authors studied a system with $N_f=2+1$ flavors for ASQTAD improved staggered fermions, at mass of $m_\pi=360$ MeV. They used the variational technique in which in addition to the standard gluonic operators they also included operators resembling glueball scattering states. There is agreement with our results for the scalar and tensor glueball masses within the bounds of statistical accuracy. The authors obtained no results for the mass of the pseudoscalar glueball. These results are shown in yellow color in Figure~\ref{fig:plot_comparison_unquenched}.

Moving on to Ref. \cite{Sun:2017ipk}, the work was carried out using $N_f=2$ flavors with the masses of clover-improved Wilson fermions corresponding to pion masses of $m_\pi = 650$ MeV and $m_\pi = 960$ MeV. Here, only gluonic operators built from Wilson loops were used in the calculation of the corresponding correlation functions. While the masses of the tensor and pseudoscalar glueballs are in agreement with our findings, there is disagreement with regards to the mass of the scalar glueball; this is true for both fermion masses the authors used. More specifically, the author's findings on the scalar glueball mass show a suppression by $ \sim 20 \% $ compared to our results. The above results for $m_\pi = 650$ MeV and $m_\pi = 960$ MeV are denoted in black and purple respectively in Figure~\ref{fig:plot_comparison_unquenched}.

Furthermore, comparing our results to those of \cite{Chen:2021dvn}, we find an agreement on the scalar and tensor glueball masses as well as an agreement within one sigma for the pseudoscalar one. In this work, the authors used gluonic operators on two $N_f=2+1$ RBC/UKQCD gauge ensembles at fermion masses corresponding to physical pion masses of $m_\pi = 140$ MeV. To enhance the statistics the authors used cluster decomposition which breaks translation invariance spontenaously. The above results are presented in yellow in Figure~\ref{fig:plot_comparison_unquenched}.

We now proceed to compare our results with data obtained on the quenched glueball spectrum. First, we consider the work in \cite{Chen:2005mg}, in which the authors used improved lattice gauge actions on anisotropic lattices. Their findings for the scalar, tensor and pseudoscalar glueball masses are all in agreement with ours.

We now take a look at the results obtained in yet another work on the quenched glueball spectrum \cite{Athenodorou:2020ani}. As evident in Figure~\ref{fig:plot_comparison_unquenched}, while the mass of the tensor glueball is in full agreement with our findings, the mass of the scalar as well as pseudoscalar agree with our data, at the level of one sigma. The results from Ref. \cite{Athenodorou:2020ani} are presented in brown colour in Figure~\ref{fig:plot_comparison_unquenched}.

 \section{Conclusions}
\label{sec:conclusions}

In this paper we have obtained the spectrum of glueball masses in the scalar, tensor and pseudo-scalar channels, $J=0^{++}$, $2^{++}$ and $0^{-+}$ respectively, using lattice configurations produced with light dynamical quarks. First, we obtained the spectrum of glueballs on configurations produced with $N_f=4$ degenerate light quarks with masses corresponding to a pion mass of $m_\pi \sim 250$ MeV. Furthermore, we investigated possible finite volume as well as discretization effects. We adopted as a reference investigation the spectrum calculated within the ensemble \texttt{cB4.06.24} for which the pion mass exhibits negligible finite volume effects. We compared our results to the data obtained using pure gauge configurations by keeping $t_0/a^2$ fixed, assuming that lattice artifacts are insignificant compared to the large statistical uncertainties accompanying these results. Strikingly, the $A_1^{++}$ spectrum appears to contain an additional (lightest) state, whose mass is between 2 and 4 pion masses, suggesting that it could be a multipion state; this is one of the most significant results of this work. Given that the first excitation level of $A_1^{++}$, for $N_f=4$, has a mass close to the ground state in pure gauge, one may hypothesise that the ground state for $A_1^{++}$ in $N_f=4$ corresponds to a decay state of the glueball. By contrast, the glueball $2^{++}$ ground state in $N_f=4$ is in good agreement with the $2^{++}$ ground state in the $SU(3)$ gauge theory and, in addition, the $0^{-+}$ ground state in $N_f=4$ is very close to the $0^{-+}$ ground state of the $SU(3)$ theory.

We attempted to shed more light on the additional state appearing in the $A_1^{++}$ channel by extracting the spectrum of the $A_1^{++}$ irreducible representation with $N_f=2+1+1$ Twisted Mass fermions for a fixed lattice spacing of $\sim 0.09$ fm and two different pion masses revealing that the mass of the $A_1^{++}$ ground state depends strongly on the pion mass, unlike the first excited state. This can be interpreted as a confirmation that the $A_1^{++}$ additional state does indeed have a large "quark content". This suggests that in future work multimeson operators should be included in the variational basis in order to identify mixings. Taken altogether our findings, in combination with data obtained from other projects, suggest that the glueball spectrum depends very little on the mass of dynamical quarks.
 
Finally, we extracted the torelon mass, and from that the associated string tension. The main outcome of this analysis is that (1) the energy of the torelon increases with its length and is consistent  with the Nambu-Goto bosonic string expression (\ref{fig:plot_nambu_goto}) and that (2) the string tension is suppressed by 50-70 \% due to the inclusion of the $N_f=4$ dynamical quarks. Given that string breaking can take place due to the inclusion of dynamical quarks, this  was perhaps unexpected.

 \begin{table}[H]
    \centering
    \begin{tabular}{c|c|c|c}
        \hline \hline
 Ensemble & ground state & $1^{\rm st }$ excited & $2^{\rm nd }$ excited \\
        \hline \hline
cB4.06.16 & 0.4257(297) & 0.6303(364)  & 0.8462(525) \\
cB4.06.24 & 0.4449(264) & 0.6445(422)  &  0.9916(743)\\
cC4.05.24 &  0.3666(213) & 0.5131(524) &   0.7210(766)  \\  
\hline
\end{tabular}
\captionof{table}{The energies in lattice units $(aE)$ for the ground, first and second excited states for the $A_1^{++}$ irreducible representation and the three $N_f=4$ ensembles.}
    \label{tab:masses_A1++}
\end{table}

 \begin{table}[H]
    \centering
    \begin{tabular}{c|c|c}
        \hline \hline
 Ensemble & ground state & $1^{\rm st }$ excited  \\
        \hline \hline
cB4.06.16 & 0.5146(118) & 0.8722(667)  \\
cB4.06.24 & - & 0.8666(622)  \\
cC4.05.24 & - & 0.7078(449)  \\  
\hline
\end{tabular}
\captionof{table}{The energies in lattice units $(aE)$ for the ground state which corresponds to the di-torelon as well as first excited state, for the $E^{++}$ irreducible representation and the three $N_f=4$ ensembles.}
    \label{tab:masses_E++}
\end{table}

 \begin{table}[H]
    \centering
    \begin{tabular}{c|c}
        \hline \hline
 Ensemble & ground state \\
        \hline \hline
cB4.06.16 & 0.8680(627) \\
cB4.06.24 & 0.8699(740) \\
cC4.05.24 & 0.6890(410)  \\  
\hline
\end{tabular}
\captionof{table}{The energies in lattice units $(aE)$ for the $T_2^{++}$ irreducible representation and the three $N_f=4$ ensembles.}
    \label{tab:masses_T2++}
\end{table}

 \begin{table}[H]
    \centering
    \begin{tabular}{c|c}
        \hline \hline
 Ensemble & ground state \\
        \hline \hline
cB4.06.16 & 0.9276(435) \\
cB4.06.24 & 0.8252(764) \\
cC4.05.24 & 0.7182(498)  \\  
\hline
\end{tabular}
\captionof{table}{The energies in lattice units $(aE)$ for the $A_1^{-+}$ irreducible representation and the three $N_f=4$ ensembles.}
    \label{tab:masses_A1-+}
\end{table}

 \begin{table}[H]
    \centering
    \begin{tabular}{c|c|c|c}
        \hline \hline
 Ensemble & ground state & $a^2 \sigma$ & $a^2 \sigma_{SU(3)}$ \\
        \hline \hline
cB4.06.16 & 0.2167(100)  & 0.01824(60) &   0.03063(32)  \\
cB4.06.24 & 0.3655(241)  & 0.01715(100) &  0.03172(43) \\
cC4.05.24 & 0.2247(183)  & 0.011365(735) & 0.02350(18) \\  
\hline
\end{tabular}
\captionof{table}{The energies in lattice units $(aE)$ for the ground state of the torelon for the three $N_f=4$ ensembles.}
    \label{tab:masses_string}
\end{table}

\section{Acknowledgements}
We express our gratitude to K.~Hadjiyannakou and G.~Spanoudes for providing access to the $N_f=4$ and $N_f=2+1+1$ configurations. AA is indebted to David Gross, Davide Vadacchino and Biagio Lucini for critical discussions over the spectrum of glueballs. MT acknowledges support in part by the Simons Collaboration on Confinement and QCD Strings. AA was supported by the Horizon 2020 European research infrastructures programme "NI4OS-Europe" with grant agreement no. 857645. Results were obtained using Cyclone High Performance Computer at The Cyprus Institute, under access with id \texttt{p014}, the Oxford Theoretical Physics cluster, the resources of computer cluster Isabella based in SRCE - University of Zagreb University Computing Centre, as well as using computing
time granted by the John von Neumann Institute for Computing (NIC) on the supercomputer JUWELS Cluster~\cite{JUWELS} at the J\"ulich Supercomputing Centre (JSC). JF received financial support by the German Research Foundation (DFG) research unit FOR5269 "Future methods for studying confined gluons in QCD", by the H2020 project PRACE 6-IP (GA No. 82376) and by the EuroCC project (GA No. 951732) funded by the Deputy Ministry of Research, Innovation and Digital Policy and the Cyprus Research and Innovation Foundation and the European High-Performance Computing Joint Undertaking (JU) under grant agreement No 951732.

\printbibliography[heading=bibintoc]

\end{document}